# Non-isothermal wetting during impact of millimeter size water drop on a flat substrate: numerical investigation and comparison with high-speed visualization experiments

**Rajneesh Bhardwaj and Daniel Attinger***

Laboratory for Microscale Transport Phenomena,

Department of Mechanical Engineering,

Columbia University, New York, NY 10027

*Corresponding author. Tel: +1-212-854-2841; fax: +1-212-854-3304;

E-mail address: da2203@columbia.edu (D. Attinger)

**Abstract**

The objective of this work is to develop and validate a numerical model to study wetting during the impact of millimeter-size drops on a flat, smooth, solid substrate under isothermal or non-isothermal conditions. A finite-element modeling is used to simulate the transient fluid dynamics and heat transfer, considering Laplace forces on the liquid-gas boundary. The Lagrangian scheme allows a very precise tracking of the free surface deformation. In this work, the numerical model is extended to account for a temperature-dependent viscosity and for dynamic wetting at the contact line. Numerical results are presented to study the influence of the kinetic wetting parameter on the wetting incipience and behavior. Our results show the influence of wetting on the spreading and the transient drop shape. Also, numerical results are compared with high-speed visualization, for cases of isothermal and non-isothermal impact. Matching between simulations and high-speed visualization allows the determination of the value of the kinetic wetting parameter. Our main finding is that warm drops spread more than cold drops because of a reduction of viscous forces, and not because of an increase of wetting.

**Nomenclature**

| | |
|---|---|
| $c$ | speed of sound [m s$^{-1}$] |
| $c_p$ | specific heat [J kg$^{-1}$K$^{-1}$] |
| $d$ | droplet diameter [m] |
| $Fr$ | Froude number [$v_0^2 d_0^{-1} g^{-1}$] |
| $g$ | gravitational acceleration [9.81 m s$^{-2}$] |
| $h$ | Planck's constant [$6.626 \times 10^{-34}$ m$^2$ kg/s] |
| $H$ | mean surface curvature [m$^{-1}$] |
| $k$ | thermal conductivity [Wm$^{-1}$K$^{-1}$] |
| $k_B$ | Boltzmann's constant [$1.3806503 \times 10^{-23}$ m$^2$ kg s$^{-2}$ K$^{-1}$] |
| $K$ | wetting parameter |
| $M$ | Mach number [$v_0 c^{-1}$] |
| $n$ | number of adsorption sites per unit area |
| $Oh$ | Ohnesorge number [$We^{0.5} Re^{-1}$] |
| $p$ | pressure [Pa] |
| $Pr$ | Prandtl number [$\mu c_{p,\,l} k_l^{-1}$] |
| $r$ | radial coordinate [m] |
| $R$ | dimensionless radial coordinate [$r d_0^{-1}$] |
| $Re$ | Reynolds number [$\rho v_0 d_0 \mu^{-1}$] |
| $S$ | spreading parameter |
| $t$ | time [s] |
| $T$ | temperature [K] |
| $u$ | radial velocity [m s$^{-1}$] |
| $v$ | axial velocity [m $s^{-1}$] |
| $v_L$ | molecular volume of liquid [$m^3$] |
| $We$ | Weber number [$\rho v_0^2 d_0 \gamma^{-1}$] |
| $z$ | axial coordinate [m] |
| $Z$ | dimensionless axial coordinate [$z d_0^{-1}$] |

**Greek letters**

| | |
|---|---|
| $\phi$ | wetting angle |
| $\gamma$ | surface energy [$Jm^{-2}$] |
| $\kappa$ | molecular displacement frequency [$s^{-1}$] |
| $\lambda$ | molecular displacement length [m] |
| $\mu$ | dynamic viscosity [Pa s] |
| $\theta$ | dimensionless temperature [$\{T - \min(T_{1,0}, T_{2,0})\}(\lvert T_{1,0}-T_{2,0}\rvert)^{-1}$] |
| $\rho$ | density [kg $m^{-3}$] |
| $\sigma$ | stress [Pa] |
| $\tau$ | dimensionless time [$t v_0 d_0^{-1}$] |

**Subscripts**

| | |
|---|---|
| 0 | initial |
| 1 | drop/droplet |
| 2 | substrate |
| adv | advancing |
| avg | average value |
| CL | contact line |
| $l$ | liquid |
| LG | liquid-gas |
| LS | liquid-solid |
| max | maximum value |
| osc | oscillation |
| r | radial direction |
| rec | receding |
| SG | solid-gas |
| w | wetting |
| z | axial direction |
| $\infty$ | ambient |

**Superscripts**

| | |
|---|---|
| 0 | equilibrium |

# 1 Introduction

The impingement of a liquid droplet on a flat substrate is relevant to technologies such as evaporative spray cooling and rapid prototyping [1, 2]. A variety of fluids are used in these processes such as water and dielectric fluids in the spray cooling case, and molten metals or polymers in the rapid prototyping case [3, 4]. In evaporative spray cooling, liquid droplets evaporating on a hot surface remove a large amount of heat through the latent heat of evaporation, on the order of 1 $MW/m^2$ [1, 5]. Rapid prototyping aims at manufacturing rapidly, with a high flexibility and without conventional casting, metallic objects that have near-net-shape, by applying precisely a controlled jet of droplets, with sizes ranging from 100 μm to 4 mm [6-8]. Jetted materials are molten metals including carbon steel, stainless steel, copper, aluminum and bronze [6], and polymers. Solder droplet jetting is a versatile, fast and economic method for producing electric connections on electronic components, where neither masks nor screens are required [3]. The fluid dynamics during droplet impact on a solid surface is a transient, viscous flow involving complex boundary conditions such as a severely deforming liquid free surface, and dynamic wetting at the liquid-solid-gas contact line (Figure 1). The physical properties that affect the process are the liquid viscosity, surface tension, density, and the wettability of the substrate surface. Heat transfer adds complexity by inducing variations of physical properties such as density, surface tension and viscosity and modifying the wetting behavior. Impact conditions including droplet diameter and impact velocity, as well as gravity influence the fate of the droplet deposition [9, 10]. In the literature, criteria are given for estimating the relative importance of the viscous, capillary and inertia forces to understand the physical mechanism involved for droplet spreading. These criteria can also be represented by regime maps as in Schiaffino and Sonin [11].

## 1.1 Theory and numerical studies

Early numerical models targeting the problem of a droplet impact on a solid surface were strongly simplified for the sake of numerical tractability, neglecting viscosity or surface tension [9, 10, 12]. In the late 90's, more accurate and realistic results were obtained with the volume-of-fluid (VOF) method [13-17]. The VOF method is an Eulerian approach, where a variable $F$ is defined to track the location of the liquid-gas interface: $F = 1$ in the '*liquid'* cells and $F = 0$ in the 'gas' cells. Cells with intermediate values of $F$ contain the free surface. For instance, Pasandideh-Fard et al. [16] simulated the impact of molten tin droplets on a steel plate, and obtained good agreement between numerical results and photographs of impacting droplets. Measured values of advancing wetting angle were used as a boundary condition for the numerical model. Recently, Pasandideh-Fard et al. [18] treated the complex three-dimensional (3D) case of oblique impact onto a flat substrate for millimeter-size drops. The predicted drop shapes agreed well with the experimental ones. An interesting contribution of this group at the University of Toronto is a criterion showing that wetting effects are negligible when [13]

$$We >> \mathrm{Re}^{1/2} \quad \textbf{(1)}$$

where $We$ and $Re$ are the Weber and Reynolds number for the droplet, respectively. Indeed, in many numerical simulations of the droplet impact problem [4, 19-23] the effects of wetting on the fluid dynamics are neglected.

An alternative to Eulerian simulation is the use of a Lagrangian approach, which was first developed by Fukai et al. [19] to investigate the droplet impact process. In a Lagrangian approach, the mesh moves with the fluid, allowing an extremely precise tracking of the free surface deformation. The modeling in [19] was made with the Finite Element method and solved the full Navier-Stokes equations. Their simulations predicted realistic features such as the formation of a propagating ring structure (due to mass accumulation) at the periphery of droplet, as well as recoiling and subsequent oscillations of the droplet. This modeling [19] was extended by Zhao et al. [20] for heat transfer to simulate the cooling of liquid metal (tin and aluminum) and water droplets. The authors found that heat transfer between an impinging microdroplet and a substrate at a different initial temperature occurs simultaneously with spreading. Subsequently, Wadvogel and Poulikakos [21] extended the model to account for solidification and modeled the impact of solder droplets on composite two-layer substrates. Their results documented the effects of impact velocity, initial droplet diameter and substrate thermophysical properties on the heat transfer and fluid dynamics of a deforming solder droplet. The numerical model predicted that heat transfer time scales were comparable to the droplet deformation scales. The numerical model used in this study is based on the same Lagrangian code as described in [4, 21, 22, 24] with the addition of a realistic treatment of wetting and temperature-dependent viscosity.

An adequate description of wetting effects during droplet deposition is important for the accuracy of the simulation. While deposition processes occurring at low Weber numbers (*We*) are entirely controlled by capillary forces [11], there is also a strong influence of wetting in the latter stages of inertia-driven impact cases. Theoretically, wetting can be described as the motion of a fluid-fluid interface (liquid-gas) over a solid surface [25]. The contact line or wetting line is the line where the three phases, solid, liquid and gas are in contact with each other. In the case of a droplet deposition with negligible inertia, the spreading parameter $S$ predicts whether a substrate is wettable or not [25]:

$$S = \gamma_{SG} - (\gamma_{LS} + \gamma_{LG}) \quad \textbf{(2)}$$

where $\gamma_{SG}$, $\gamma_{LS}$ and $\gamma_{LG}$ are the interfacial energies between solid and gas, liquid and solid and liquid and gas, respectively. For $S > 0$, the liquid spreads into a nanometer-thin film, a phenomenon called total wetting. For $S < 0$, the drop does not spread completely but reaches an equilibrium shape as a spherical cap with equilibrium wetting angle $\phi$: this is known as partial wetting. In the latter case, the Young equation relates wetting angle and the surface energies as shown in Figure 2:

$$\gamma_{LG} \cos\phi = \gamma_{SG} - \gamma_{LS} \quad \textbf{(3)}$$

While Young's relation describes an equilibrium configuration, any tension (force per unit length) imbalance in Figure 2 induces a motion of the wetting line, a phenomenon called dynamic wetting. Realistic models for dynamic wetting involve an accurate description of the fluid flow and molecular events in the vicinity of the wetting line, in a treatment compatible with the bulk droplet flow. Two classes of models are currently available: a first category of models follow the hydrodynamic approach, allowing slip between the liquid and the solid at the wetting line to avoid singularities in mass and momentum conservation [26]. This hydrodynamic approach does not consider molecular mechanisms at the wetting line, but rather explains the evolution of the apparent dynamic wetting angle by hydrodynamic bending of the free surface at sub-microscopic distances of the wetting line [27]. The apparent dynamic wetting angle is defined as the macroscopic wetting angle that can be seen through a low-power microscope, implying that the effective wetting angle, if measurable at molecular scale, would differ substantially from the apparent wetting angle. A typical relationship between the wetting angles $\phi_{adv}$ and $\phi_{rec}$ and the wetting line speed [28] is given in Figure 3, showing that the equilibrium wetting angle can exhibit hysteresis when the contact line is not moving. Average values of the advancing and receding wetting angles can be determined experimentally and used as an input to the numerical model. Pasandideh-Fard et al. [13] accounted for wetting effects in their VOF model using this technique. Their numerical results showed a better agreement with experimental observations when the advancing wetting angle was obtained from photographs and input in the code, rather than when the wetting angle was set *a priori*.

A second category of models are the molecular kinetic models: they are based on microscopic jumps of molecules at the wetting line [29-31]. The molecular-kinetic modeling used in the present study expresses the movement of the wetting line as a statistical process involving adsorption jumps of a myriad of individual molecules [29-31]. Molecules at the wetting line cross energy barriers arising through interactions between fluid/solid and fluid/fluid molecules. In this framework, a stationary wetting line is nothing else than an interface with zero net molecular flow rate in either advancing or receding direction. Any driving potential, typically a stress induced by a departure from the equilibrium wetting angle, will alter the energy barriers asymmetrically and favor the movement in a specific direction, as described by Blake and de Coninck [32]. The expression relating the wetting line speed $u_{CL}$ to the dynamic wetting angle $\phi$ is given in section 2.1.2.2.

## 1.2 Experimental studies

High-speed photography is a primary source of information about droplet impact and spreading. For instance, the impact and spreading dynamics of an n-heptane droplet, with an initial diameter of about 1.5 mm, on a smooth stainless steel surface was visualized by Chandra and Avedisian [33]. After 2.6 ms of impact, the droplet assumed the shape of a flattened disc. The liquid kept spreading to reach a maximum diameter of 6 mm at approximately 20 ms after impact. In their work, they used single-shot flash photographic technique, in which a flash light with an 8 microsecond shot duration freezes the droplet shape at a given instant of its impact. Then, by varying the delay between the droplet impact and the illumination, they photographed successive stages of the deposition process, so that the entire process could be pieced together from individual images [33]. Using flash videography with 150 ns light pulses, Attinger et al [34] visualized the impact, spreading, oscillations and solidification of a molten solder microdroplet, a process occurring within less than 300 microsecond. The dynamic interaction between the oscillation in the liquid region and the rapid advance of the solidification front was visualized, quantified and presented. Recently, flash videography technique has been successfully used to record the impact of a micrometer size water drop on a glass substrate [35] and polycarbonate substrate [36]. The impact of the drop on a solid substrate can also result in splashing and rebound of the drop. Rioboo et al [37] recorded the impact of water, ethanol, liquid alloys, different mixtures of glycerin and water and silicone oil on solid surfaces with different roughness and wettability. They observed different outcomes such as deposition, splashing, receding and rebound during the impact of the drop. A detailed review of splashing, receding and rebound of drops is presented in [38].

Key parameters describing the droplet spreading process can be subsequently measured from the recorded pictures, such as the droplet diameter, height and wetting angle. Known limitations of visualization methods are the temporal resolution, limited by frame rate, and spatial resolution, limited by the imaging system (lens aberrations, viewing angles, lighting, shadows at high wetting angles), the resolution of image recording media, and the image reconstruction process. Also, images are two-dimensional, while the deformation is three-dimensional: some portions of the free surface can therefore be hidden by others. Finally, the fluid dynamics inside the drop, as well as other transport phenomena, is extremely difficult to visualize and measure, or even impossible in the case of a non-transparent liquid.

The primary motivation of the work presented in this article is a better understanding of the wetting phenomenon associated with the impact of a droplet on a flat substrate in the presence of heat transfer. Technically, the aim is to determine realistic parameters in a kinetic wetting modeling. First (section 2), an existing Finite-Element model for non-isothermal droplet impact is used and expanded to account for wetting and temperature-dependent viscosity. Second (section 5.1.1), the sensitivity of the model to the wetting parameter is studied and described. Third (section 5.2), high-speed visualization experiments are performed and matched with the simulations to identify the actual value of the dynamic wetting parameter. The droplet used in this work has an initial diameter of 3.2 mm, impact velocity of 0.4 m/s. The initial droplet temperature is varied between 25$^{o}$C (ambient) and 60$^{o}$C, while the substrate is at ambient temperature (25$^{o}$C). This corresponds to values of Weber and Froude number of 7.1 and 5.1, respectively. The range of the value of Reynolds number for these conditions is from 1400 to 2800.

# 2 Mathematical model

The mathematical model is based on the unsteady solution of Navier-Stokes and energy equation in an axisymmetric geometry. All equations are expressed in a Lagrangian framework [19, 21].

## 2.1 Fluid dynamics

The radial and axial components of the momentum equation are considered along with the continuity equation. An artificial compressibility method is employed to transform the continuity equation into a pressure evolution equation. This method assumes a fluid flow that is slightly compressible, whereby the speed of sound is large, but not infinite [19]. A Mach number (*M*) of 0.001 is used for all simulations in this work. The flow inside the droplet is laminar and variations of viscosity with temperature are taken into account. All other thermophysical properties are assumed to be constant with respect to temperature. The dimensional form of the fluid dynamics equations are [9]:

*Mass conservation:*

$$\frac{1}{c^2}\frac{Dp}{Dt}+\rho\left(\frac{1}{r}\frac{\partial}{\partial r}(ru)+\frac{\partial v}{\partial z}\right)=0 \qquad \mathbf{4}$$

where $c$, $p$, $t$, $\rho$, $r$, $z$, $u$, $v$, are speed of sound, gage pressure, time, density of liquid, radial distance, axial distance, radial velocity and axial velocity, respectively.

*Momentum conservation in radial direction:*

$$\rho\frac{Du}{Dt}-\frac{1}{r}\frac{\partial}{\partial r}\left(r\sigma_{rr}\right)-\frac{\partial\sigma_{rz}}{\partial z}+\frac{1}{r}\sigma_{\theta\theta}=0 \qquad \mathbf{5}$$

*Momentum conservation in axial direction:*

$$\frac{Dv}{Dt}-\frac{1}{r}\frac{\partial}{\partial r}\left(r\sigma_{rz}\right)-\frac{\partial\sigma_{zz}}{\partial z}-\rho g=0 \qquad \mathbf{6}$$

where $\sigma_{ij}$ are the corresponding components of the stress tensor, which are defined as follows:

$$\sigma_{rr} = -p + 2\mu \frac{\partial u}{\partial r}; \sigma_{zz} = -p + 2\mu \frac{\partial v}{\partial z};$$

**7**

$$\sigma_{\theta\theta} = -p + 2\mu \frac{u}{r}; \sigma_{rz} = \sigma_{zr} = \mu\left(\frac{\partial u}{\partial z} + \frac{\partial v}{\partial r}\right)$$

The above governing equations are non-dimensionalized according to the procedure in [21], where their non-dimensional expressions are given.

### 2.1.1 Initial conditions

The initial conditions of the problem are as follows (Figure 4):

$$t = 0; u = 0; v = -v_0; p = \frac{4\gamma}{d_0} \qquad \mathbf{8}$$

where $v_0$ is the impact velocity, $d_0$ is the initial diameter of the droplet, $\gamma$ is surface tension of the liquid and $p$ is the gage pressure.

### 2.1.2 Boundary conditions

#### *2.1.2.1 At free surface and along z-axis*

In the derivation of the boundary condition at the free surface, a balance is considered between forces due to pressure, viscosity and surface tension [39]. The free surface boundary conditions are given by [9]:

$$\sigma_{rr} n_r + \sigma_{rz} n_z = -2H\gamma n_r \qquad \mathbf{9}$$

$$\sigma_{zr} n_r + \sigma_{zz} n_z = -2H\gamma n_z \qquad \mathbf{10}$$

where $n_r$ and $n_z$ are the radial and axial components of the outward unit normal to the surface and $H$ is the mean curvature. Symmetry along the z-axis implies:

$$r = 0; u = 0; \frac{\partial v}{\partial r} = 0 \qquad \textbf{11}$$

*2.1.2.2 Along r-axis and at wetting line*

The no-slip boundary condition below is applied at the $z = 0$ plane, except at the wetting point

$$z = 0; r \neq r_{CL}; u = 0; v = 0 \qquad \textbf{12}$$

The boundary condition at the wetting line ($r_{CL}$, 0) is a critical condition for having a good agreement with experiments. Many researchers have demonstrated that the traditional no-slip boundary condition fails in the vicinity of the wetting line causing an infinite stress in the region [28]. To circumvent this problem, slip models have been postulated whereby a small finite region of fluid at the wetting line is assumed to slide along the impacted surface. To this end, following the discussion in Bach and Hassager [40] two types of boundary conditions must be specified at each location on the fluid boundary. Two essential (velocity) conditions must be specified on a no-slip surface. If slip is considered, one essential and one natural (force) condition are required. One way to do so is therefore to apply the essential boundary condition $v = 0$ and the natural condition $\sigma_{rr} n_r + \sigma_{rz} n_z = -2H\gamma n_r$ at the contact point ($r_{CL}$, 0). This formulation, which is adopted from the work of Bach and Hassenger [40], does not specify a wetting angle or a dynamic velocity at the wetting line, but merely applies an interfacial force extrapolated from the Laplace force along the free surface for the sake of avoiding an infinite stress at the wetting line. This level of modeling is consistent with the assumption of a constant coefficient of surface tension in the free surface force balance given by Equations 9 and 10. Since this approach typically underestimates the spreading magnitude of the cases computed here, we call this approach the low-wetting approach.

Alternatively, a dynamic velocity $u_{CL}$ at the wetting line can be formulated according to kinetic theory and coupled to the essential boundary condition $v$ = 0. The kinetic model for the horizontal velocity $u_{CL}$ at the wetting line from Blake and de Coninck [32] is implemented. This model describes wetting as a dynamic adsorption/desorption process of liquid molecules to the surface. For the wetting line to move, work must be done to overcome the energy barriers to molecular displacement in the preferred direction. This work is provided by the out-of-balance surface tension force ($=\gamma_{LV}(\cos\phi^0 - \cos\phi)$) where $\gamma_{LV}$ is the liquid-vapor surface energy (J/m$^2$), $\phi^0$ is the equilibrium wetting angle and $\phi$ is the dynamic wetting angle. The wetting line velocity ($u_{CL}$) during the wetting of liquid on a solid surface is given in [32] :

$$u_{CL} = \frac{2\kappa_s^0 h\lambda}{\mu v_L}\sinh\left[\frac{\gamma_{LV}}{2nk_BT}(\cos\phi^0 - \cos\phi)\right] = \frac{2K_w\lambda}{\mu}\sinh\left[\frac{\gamma_{LV}}{2nk_BT}(\cos\phi^0 - \mathrm{c}\right. \qquad \mathbf{13}$$

where $\lambda$ is molecular displacement length (O(angstrom)), $n$ is the number of adsorption sites per unit area, $k_B$ is the Boltzmann's constant ( = 1.3806503 × 10$^{-23}$ m$^2$ kg s$^{-2}$ K$^{-1}$), $T$ is the absolute temperature (K) and $\kappa_s^0$ is a constant involving the molecular equilibrium displacement frequency (s$^{-1}$), $\kappa_w^0$ [32]:

$$\kappa_w^0 = \kappa_s^0\left(\frac{h}{\mu v_L}\right) \qquad \mathbf{14}$$

In the above expression, $h$ is Planck's constant (= 6.626 × 10$^{-34}$ m$^2$ kg s$^{-1}$), $\mu$ is the temperature dependent viscosity and $v_L$ is molecular volume of liquid [m$^3$]. In this study, $K_w = \kappa_s^0 h/v_L$ is used as the parameter controlling the wetting speed and its value is found by comparing numerical results with experimental ones, assuming that $\lambda = 2 \times 10^{-10}$ m and $n = \lambda^{-2}$ [41]. More details on the comparison between experimental and numerical results are given in section 5.2. In our simulations, the equilibrium wetting angle is taken as zero ($\phi^0 = 0^o$), because water wets glass, and the dynamic wetting angle $\phi$ is obtained from the simulation.

For either the low-wetting and the kinetic approach, the boundary conditions at the wetting line are expressed as:

$$\text{At } z = 0; r = r_c; v = 0;\ \sigma_{rr} n_r + \sigma_{rz} n_z = -2H\gamma n_r \text{ (Low-wetting approach)} \quad \mathbf{15}$$

$$\text{At } z = 0; r = r_{CL}; v = 0; u = u_{CL}(\phi) \text{ (Equation 13, Kinetic approach)}$$

## 2.2 Heat transfer

The conduction and convection heat transfer is solved in both the droplet and the substrate, according to the formulation in [21]. At the droplet free surface, convection with air and radiation heat transfer are neglected. During complete cooling of 3.2 mm drop, the convection and radiation losses are around 1.71% and 0.87% of initial thermal energy of drop, respectively [42]. The energy conservation equation for the droplet and substrate is given by ($i$ = 1 for droplet and $i$ = 2 for substrate) [10]:

$$\rho_i c_{p,i} \frac{DT_i}{Dt} - \frac{1}{r}\frac{\partial}{\partial r}\left(k_i r \frac{\partial T_i}{\partial r}\right) - \frac{\partial}{\partial z}\left(k_i \frac{\partial T_i}{\partial z}\right) = 0 \quad \mathbf{16}$$

where $c_{p,i}$ denotes specific heat, $k_i$ is thermal conductivity and $T_i$ is temperature. The remaining quantities were defined earlier in the development of the fluid dynamics model. Note that this model accounts for both conduction and convection in the drop since the Lagrangian approach suppresses the need for explicit convection terms. A variable viscosity is incorporated through the local Reynolds number $\mathrm{Re} = \dfrac{\rho v_0 d_0}{\mu(T)}$ for each element in the computational domain. Note that the viscosity is calculated as a function of temperature (section 3.2). The above governing equations are non-dimensionalized according to the procedure in [21], where their non-dimensional expressions are given.

#### 2.2.1 Initial and boundary conditions

Initial conditions in the droplet and substrate regions are:

$$T_1(r,z,0) = T_{1,0}\,;\; T_2(r,z,0) = T_{2,0} \qquad \textbf{17}$$

Heat transfer from the exposed droplet and substrate surfaces is neglected, such that the following condition applies to the boundaries of both regions ($i$ = 1 for droplet and $i$ = 2 for substrate):

$$\frac{\partial T_i}{\partial R} n_r + \frac{\partial T_i}{\partial Z} n_z = 0 \qquad \textbf{18}$$

A perfect thermal contact is assumed at the interface between the solid and the liquid [19].

## 3 Numerical scheme

The computational domain is discretized as a mesh of triangular elements and the numerical model is solved using a Galerkin finite element method. Linear shape functions are used for velocity and pressure. An implicit method is utilized for the integration of the fluid dynamics equations in time [40], while a Crank-Nicholson scheme is used for the energy equation. Details of the algorithm are given in [21]. A free and robust mesh generator, *Mesh2D,* is used in the present modeling [43]. A typical mesh used in computational domain is shown in Figure 5a.

### 3.1 Mesh size and time step independence

Although the numerical code used in this study has been validated in the past for related cases [20, 21, 24], it is important to verify that the implementation of a different mesh generator, a new wetting model and a different parameter domain do not affect the complex data structure nor perturb the solution. The grid and time step independence are therefore examined in terms of the maximum height ($Z_{max}$) and wetted radius ($R_w$) of droplet, as shown in Figure 5b. This study is carried out for a 3.2 mm water droplet impacting a flat surface at 0.25 m/s under isothermal conditions. This corresponds to $Re$ = 814 and $We$ = 2.74. Kinetic wetting is taken into account with the wetting parameter, $K_w = 2.5 \times 10^6$ Pa. The grid influence is considered for five increasing number of nodes in the droplet: 750, 850, 950, 1050 and 1200 with a time step of $5 \times 10^{-4}$ in each case (Figure 6a). The substrate nodes are 3200 with adaptive mesh in each case. The substrate size is a square of 4 × 4 non-dimensional size. Results for 1050 and 1200 nodes are so close that 1050 droplet nodes are chosen for the final simulations (Figure 6a). The time step influence is considered for time steps of 5, 4, and $3 \times 10^{-4}$ for 1050 nodes in each case, with the results shown in Figure 6b. Time steps $1 \times 10^{-3}$ and $2 \times 10^{-3}$ sometimes cause excessive deformation at the wetting line so that the code crashes and has to be manually restarted. The results for all three-time steps are similar so that a time step $5 \times 10^{-4}$ is chosen (Figure 6b). A typical fluid dynamics simulation until complete rest of drop i.e. when the kinetic energy of drop is 1% of the initial kinetic energy takes approximately 10 hour of CPU time on a 3.2 GHz Intel-Xeon machine with 1 GB of RAM.

## 3.2 Thermophysical properties

The thermophysical properties used for the simulations are given in Table 1. In the mathematical model, the temperature-dependence of viscosity is considered. For this purpose, the following regression relation is obtained from data in [44]:

$$\mu = -0.0026\,T^3 + 0.5874\,T^2 - 47.598\,T + 1763.4 \qquad (19)$$

where $\mu$ is the viscosity of water in μPa-s and $T$ is the temperature in degree C. Within the expected range of temperature inside the drop (25°C to 60°C) in this study, the decrease in value of viscosity and surface tension are around 48% and 8%, respectively from its value at 25°C [44]. Marangoni effect is not considered. The reason for this is as follows: it is very likely that in the case of the impact of the water droplet, unintentional surfactant contaminants have reduced the Marangoni flow [45, 46]. According to analytical calculation of Hu and Larson [45], a tiny concentration of contaminants ($1 \times 10^9$ molecules for 1000 micrometer wetted diameter water drop) can greatly reduce Marangoni stresses. The evaporation of the drop in the ambient is also neglected. It can be justified as follows: the total evaporation time for a 3.2 mm water drop on fused silica given by the analytical formula of Popov [47] is 50 minutes (assuming 8 mm wetted diameter of the drop after it comes to rest after impact, 20° contact angle and 40% relative humidity). The evaporation time is 4 orders of magnitude larger than the time in which the drop comes to rest (~ 200 ms) after the impact on the substrate. Thus, evaporation during the impact of the drop is neglected. However, in some cases such as the thin-film flows over heated surfaces [48, 49] or boiling in microchannels [50], evaporation and Marangoni stresses are important and should be accounted for [51]. The variation of liquid thermal conductivity is also neglected in the numerical model since the increase in thermal conductivity of water at temperatures from 25$^{o}$C to 60$^{o}$C is about 8% [44].

# 4 Experimental setup

The configuration of the experimental setup is shown in Figure 7. A 3.2 mm droplet is generated via dripping from an Eppendorf 200 μL polypropylene nozzle (520 μm internal diameter). The liquid flow is controlled using a solenoid valve and the height of a reservoir. The height of the nozzle above the fused-silica substrate (prism) can be adjusted to vary the impact velocity from 0.1 to 1 m/s. This prism provides a smooth, flat surface for droplet impingement. The prism is carefully cleaned using isopropanol and dried between each impact. The initial droplet temperature can be controlled from 25$^{o}$C (ambient temperature) to 80$^{o}$C using a Ni-Cr heating wire coiled around the nozzle. This initial temperature is measured with a K-type thermocouple (Omega Inc, 0.25 mm diameter) protruding into the dripping drop. This thermocouple is connected to a Keithley digitizer, with a time resolution is 0.1 s, and a temperature resolution is ±0.5$^{o}$C. High-speed visualizations of the droplet impact are performed using a 1.3 Megapixel Pixelink digital firewire camera (Pixelink, PLA 741) with a zoom objective. The chosen magnification corresponds to 75 μm per pixel, which implies a dimensional error of ± 75 μm. Both the exposure time and image size of the camera are adjustable, allowing for the maximization of firewire bandwidth. Typically, a frame rate of 3000 fps is achieved using an image size of 120 × 50 pixels. Experimental images are analyzed using the image analysis software Scion. The cooling of the drop during free fall is negligible: according to the analytical model in [52] for transient, one-dimensional heat conduction in sphere (drop) the maximum decrease of temperature at the droplet free surface is less than 1$^{o}$C for the range of initial temperatures considered in this study .

# 5 Results and discussion

## 5.1 Numerical results

Results are presented in this section for a water drop impacting on a fused silica substrate. The drop diameter ($d_0$) is 3.2 mm, the impact velocity $v_0$ = 0.4 m/s and the drop initial temperature is varied between the ambient temperature (25$^{o}$C) and 60$^{o}$C. This corresponds to values of Weber and Froude number of 7.1 and 5.1, respectively. The Reynolds number for isothermal impact is 1418 while the initial values (before the impact of the drop) of Reynolds number for non-isothermal impacts: $T_{1,0}$ = 45$^{o}$C and $T_{1,0}$ = 65$^{o}$C are 2223 and 2771, respectively. The simulation cases are chosen so that they correspond to experimentally obtainable values using dripping from a small-diameter tube, placed at a defined height above the substrate. The reasons to select these values of diameter and impact velocity for the numerical study are mainly related to the experimental and physical characteristics of the problems:

- Drops with diameters smaller than 1 mm are difficult to generate by dripping, because of the negligible influence of gravity.
- Impact velocity lower than 0.1 m/s are difficult to set experimentally because the velocity is controlled by the height between the substrate and the drop release location.
- Drops with diameters larger than 4 mm lose sphericity because of their high Weber number.
- Drop splashing is difficult to simulate with our finite-element method, because interface breakup requires corresponding changes in the code datastructure. Splashing occurs when Oh.$Re^{1.25}$ > 57.7 [53] with Oh is Ohnesorge number. With the drop size considered here (3.2 mm), splashing would occur at velocities greater than 1.1 m/s.

A total of 5 simulations are performed as listed in Table 2.

### 5.1.1 Influence of wetting

As discussed in section 2.1.2.2, the kinetic wetting model given by Blake and de Coninck [32] is used in the simulations (Equation 13). The magnitude of wetting is controlled through the wetting parameter value, $K_w$. In this section, the effect of wetting is studied by varying numerically the wetting parameter $K_w$. First, isothermal impact of water drop with diameter and impact velocity of 3.2 mm and 0.40 m/s, respectively ($Re$ = 1418, $We$ = 7.1) is considered. Figure 8 shows the droplet deformation for the low wetting model (left), wetting ($K_w = 2 \times 10^6$ Pa, middle) and larger wetting case ($K_w = 2 \times 10^7$ Pa, right). These three simulations correspond to # 1 (low wetting), # 2 ($K_w = 2 \times 10^6$ Pa) and # 3 ($K_w = 2 \times 10^7$ Pa) in Table 2. By comparing the droplet deformations in all three cases (Figure 8a, b), the effect of increasing wetting can be explained. When there is low wetting, the droplet spreads under the influence of inertial forces ($t$ = 6.4 ms). Surface forces slow the spreading until the drop assumes a doughnut shape ($t$ = 7.2 ms). When a significant amount of wetting comes into play (for the two columns on the right), wetting works hands in hands with inertia to increase the spreading of the droplet. For instance, increased spreading due to wetting is visible as early as $t$ = 3.2 and 0.8 ms for the respective middle and right columns in Figure 8. A visual inspection of the droplet deformation near the contact line confirms this fact. At $t$ = 7.2 ms, the wetted area is 20% larger in the case with $K_w = 2 \times 10^7$ Pa than in the low-wetting case. A mathematical reason for that is that the wetting line velocity is proportional to $K_w$. A comparison of non-dimensional wetted splat radius and maximum height of splat with respect to time is given for all three cases in Figure 9a. A larger value of the wetting line velocity therefore results in an earlier wetting and larger spreading. A comparison of the evolution of contact angle for three cases is shown in Figure 9b. In the 'low wetting' case, the contact angle remains larger than $155^{o}$ for the first 3 ms of the spreading. For $K_w = 2 \times 10^6$ Pa and $K_w = 2 \times 10^7$ Pa, a sudden drop in the contact angle is observed at 1.0 and 0.1 ms, respectively. This sudden drop in contact angle shows the transition to wetting in both cases. This confirms that both the initiation of wetting and its speed depend on the value of wetting parameter. As shown in section 5.2, the best agreement between experiments and simulations is obtained for $K_w = 2 \times 10^6$ Pa.

### 5.1.2 Coupling of fluid dynamics and heat transfer

Temperature influences fluid dynamics by varying viscosity, and this effect is investigated numerically as follows (Figure 10). The spreading of a water droplet on a 25°C fused silica solid surface is considered for three initial droplet temperatures: 25°C (isothermal impact), 45°C and 60°C. These three simulations are numbered as # 2, 4 and 5 in Table 2. The value of Weber number for all three cases is 7.1. The Reynolds number for the isothermal impact is 1418 while the initial values (before the impact of the drop) of Reynolds number for the non-isothermal impacts: $T_{1,0}$ = 45°C and $T_{1,0}$ = 60°C are 2223 and 2771, respectively. The kinetic wetting parameter is $K_w = 2 \times 10^6$ Pa for all three cases. Figure 10 shows the evolution of wetted radius and maximum height of the drop for three cases mentioned. The largest spreading is observed when the initial drop temperature is 60°C while the smallest corresponds to the 25°C case. This corresponds to an increase of 17% in the value of wetted radius. The reason is that warming the drop decreases its viscosity and correspondingly the ratio of inertial to dissipative forces (Reynolds number). This results in a larger spreading of the drop. The change in wetted radius is larger when the temperature increases from 25°C to 45°C, than when the temperature is increased from 45°C to 60°C. This probably arises because the variation of viscosity of water from 25°C to 45°C is around 30% while the variation from 45°C to 60°C is around 22%. Another interesting change due to the temperature increase is that the oscillations of the drop height are smaller and damp faster, because the larger spreading associated with a higher temperature prevents large height oscillations as seen in Figure 10. Interestingly, the simulations show that the drop temperature at the wetting point drops to the substrate temperature very early in the spreading process.

## 5.2 Comparison of experimental and numerical results

In this section, high-speed visualization results are compared with the corresponding simulations for a 3.2 mm water drop impacting on a flat fused silica substrate with velocity 0.40 m/s. The initial temperature of the drop is ambient temperature (25°C), 45°C or 60°C while the substrate is at ambient temperature initially in all cases.

### 5.2.1 Comparison of isothermal impact

Figure 11 compares a sequence of high-speed visualization images (left) with simulated instantaneous streamlines and free surface (right) for the impact of a 3.2 mm diameter water droplet at 0.4 m/s onto a fused silica substrate at the same temperature of 25°C. The Reynolds and Weber numbers are respectively 1418 and 7.1. Care is taken to have the same scaling for both images. The visualization results do not show significant asymmetry during impact. The substrate level is indicated with a straight horizontal line right of the simulation images. The parameters used in the numerical simulation correspond to the case presented in section 5.1.1 (simulation # 2 in Table 2 and the middle column of Figure 8). Several features predicted by the numerical simulations are confirmed by the high-speed visualization. During the initial spreading stage ($t$ < 6 ms), the motion of the drop is mostly due to inertial and viscous forces, as shown by the strong deformations of the free surface and the flow pattern in the drop. This extensional impact flow reverses at 12 ms. Between $t$ = 12 and 18 ms, surface tension causes the drop to recoil and the principal flow direction is from the edges towards the drop center, inducing an increase of the drop height at its center and a film-like pattern at its periphery. For times larger than 18 ms, the flow reverses again as the height of the droplet decreases ($t \sim 28$ ms). After several of these oscillations, the drop comes to rest at about 200 ms. The agreement between numerical and experimental droplet shape is good, in terms of initial shape, oscillations frequency and amplitude. The major difference between the visualization and the simulations occurs at $t$ = 8 ms, where it seems that the drop height at the center of the picture is lower in the simulation than in the visualization. A reason for that, however, comes from the fact that the visualization is a side image, while the simulation is a radial cut, which explains the difference. Finally, a comparison of numerical and experimental values for wetted radius and maximum height of the drop is shown in Figure 12. The agreement for dynamic behavior of the radius and droplet maximum height is good, both in magnitude and oscillation frequency, with a final value of the numerical wetted radius about 6% larger than the value given by experiments.

### 5.2.2 Comparison of non-isothermal impact

The comparisons of evolution of wetted radius and maximum height of the drop for numerical and experimental cases are shown for two non-isothermal cases (initial drop temperature is 45$^{o}$C or 60$^{o}$C) in Figure 13 and Figure 14, respectively. The impact velocity of the drop is 0.40 m/s, which corresponds to Weber and Froude number of respectively 7.1 and 5.1. The initial values (before the impact of the drop) of the Reynolds numbers are 2223 and 2771 for initial temperature of the drop 45$^{o}$C and 60$^{o}$C, respectively. The physical parameters used in these two cases correspond to simulations # 4 and 5 in Table 2. The variation of viscosity with temperature as well as kinetic wetting is considered in the numerical model. Figure 13 shows the evolution of wetted radius and maximum height of the drop for an initial drop temperature of 45°C. Both experimental and numerical curves show that the splat radius increases with time until it attains a constant value. The experimental and numerical curves for maximum height show oscillations, which can be attributed to the competition between inertial and surface forces (as explained in section 5.2.1). After several droplet oscillations, the drop comes to rest at about 120 ms. The agreement between numerical and experimental droplet oscillation frequency is good until the second occurrence of recoiling. The latter shifting oscillation frequency might come from the variation of the surface tension (not taken into account in the model) [54], or from the non-spherical shape of the drop before impact. The final value of numerical wetted radius (when drop comes to rest) is around 2% larger than the value given by experiments. Also, compared to the isothermal impact, the drop spreads about 20% more in this non-isothermal case (initial drop temperature 45°C), which is due to the viscosity decrease due to temperature. The amplitude of oscillations of maximum droplet height is around 10% smaller than in case of isothermal impact for the similar reason. Finally, Figure 14 presents the case of a 60°C droplet impacting a 25°C substrate. Both the droplet oscillation frequency and final wetted radius are in very good agreement between the simulations and the experiments (Figure 14). The oscillation period $t_{osc}$ [s] of the drop can also be quantified by the rough analytical estimate given in Schiaffino and Sonin [11]: $t_{osc} = \sqrt{\rho d_0^3 / \gamma}$ where $\rho$, $d_0$ and $\gamma$ are density, initial diameter and surface tension of the drop, respectively. The measured periods of oscillation for three cases ($T_{1,0}$ = 25°C, 45°C or 60°C) shown in Figure 12, 13 and 14 are respectively around 20, 25 and 25 ms. The analytical relation above gives values very close to the experimental ones, in the range 21 – 22 ms. The best agreement between simulations and experiments was obtained with the same wetting parameter $K_w = 2 \times 10^6$ Pa for all three droplet initial temperatures. It is important to mention that the wetting line temperature in all the cases considered in this paper reaches within 1 millisecond a value close to the ambient temperature (see Figure 10). In other words, although the wetting line velocity depends on temperature through the viscosity $\mu$ and wetting parameter $K_w$ (Equation 13), most of the spreading occurs with a wetting line cooled to the substrate temperature. Therefore, it appears that the additional spreading observed when a heated drop is deposited on a substrate comes mainly from a reduction in viscosity in the bulk fluid, not from a wetting enhancement.

## 6 Conclusions

A numerical investigation of the fluid mechanics for a millimeter size water droplet impacting on a fused silica substrate has been performed. Dynamic wetting and the temperature dependence of viscosity have been implemented in a numerical model. The capability of the numerical model used is illustrated by five simulations involving various values of droplet initial temperature and wetting parameters for millimeter size drop. In particular, the effect of temperature variation of viscosity and influence of the magnitude of the kinetic wetting parameter are assessed. Instantaneous streamlines in our numerical results show how wetting increases the spreading of the droplet. Also, comparisons of numerical results with experimental results using high-speed visualization are performed. Numerical and experimental results are found to be in very good agreement in terms of droplet shapes, spreading and oscillation frequency. The temperature at the wetting line is found to reach the substrate temperature at a very early stage of the spreading, a finding that indicates that the larger spreading observed with heated drops is due to a reduction of viscous forces in the bulk of the liquid, and not to an increase of wetting kinetics.

## 7 Acknowledgements

The authors gratefully acknowledge financial support for this work from the Chemical Transport Systems Division of the US National Science Foundation through grant 0622849.

# 9 Tables

Table 1: Thermophysical properties used in the simulations [44, 55, 56]

| Droplet | Density ($kgm^{-3}$) (at 25$^oC$) | Thermal conductivity ($Wm^{-1}K^{-1}$) (at 25$^oC$) | Specific heat ($Jkg^{-1}K^{-1}$) (at 25$^oC$) | Viscosity (Pa-s) (at 25$^oC$) | Surface energy ($Jm^{-2}$) (at 25$^oC$) | Initial dimension-less temperature | Thermal diffusivity ($m^2 s^{-1}$) |
|---|---|---|---|---|---|---|---|
| Water | 997 | 0.607 | 4180 | $9.0 \times 10^{-4}$ | $7.2 \times 10^{-2}$ | 1.0 | $1.46 \times 10^{-7}$ |
| Substrate | | | | | | | |
| Fused silica | 2200 | 1.38 | 740 | - | - | 0.0 | $8.48 \times 10^{-7}$ |

Table 2: Cases considered*

| Simulation # | $d_0$ [mm] | $v_0$ [m/s] | *Re* | *We* | *Fr* | $T_{1,0}$ [°C] | $K_w$ [Pa] | $\mu = f(T)$ |
|---|---|---|---|---|---|---|---|---|
| 1 | 3.2 | 0.40 | 1418 | 7.1 | 5.1 | 25 | 0** | - |
| 2 | 3.2 | 0.40 | 1418 | 7.1 | 5.1 | 25 | $2 \times 10^6$ | - |
| 3 | 3.2 | 0.40 | 1418 | 7.1 | 5.1 | 25 | $2 \times 10^7$ | - |
| 4 | 3.2 | 0.40 | 2223 (initial value) | 7.1 | 5.1 | 45 | $2 \times 10^6$ | Yes |
| 5 | 3.2 | 0.40 | 2771 (initial value) | 7.1 | 5.1 | 60 | $2 \times 10^6$ | Yes |

*Initial substrate temperature ($T_{2,0}$) is 25°C in all simulations
**Corresponds to low wetting model (section 2.1.2.2)

## 10 Figure captions

Figure 1: Definition of problem to be studied

Figure 2: Surface forces at the wetting line

Figure 3: Relationship between the advancing and receding wetting angles, and the wetting line velocity. From [27].

Figure 4: Initial conditions of the problem to be studied

Figure 5: (a) A typical mesh in computational domain (b) Definition of the maximum height ($Z_{max}$) and wetted radius of droplet ($R_w$)

Figure 6: (a) Grid independence study: Variation of wetted radius and maximum height of droplet with time for different numbers of nodes in the droplet (b) Time-step independence study: Variation of wetted radius and maximum height of droplet with time for different time steps.

Figure 7: Experimental setup

Figure 8a: Fluid dynamics during isothermal impact of a water drop on fused silica substrate (0 to 3.2 ms) for three different treatment of wetting (low wetting model, wetting with $K_w = 2\times10^6$ Pa and with $K_w = 2\times10^7$ Pa ). Diameter of drop and impact velocity are 3.2 mm and 0.40 m/s respectively ($Re$ = 1418, $We$ = 7.1) in each case. Initial temperature of drop as well as substrate is 25°C in each case (isothermal impact).

Figure 8b: continuation of (a) for latter times

Figure 9: (a) Time evolution of dimensionless wetted radius and maximum height of splat for three different treatments of wetting (low wetting model, wetting with $K_w = 2\times10^6$ Pa and with $K_w = 2\times10^7$ Pa). The drop diameter and impact velocity in each case are 3.2 mm and 0.4 m/s,

respectively. (b) Time evolution of contact angle for the three different treatments of wetting. Initial temperature of drop as well as substrate is 25$^o$C in each case (isothermal impact).

Figure 10: Time evolution of dimensionless wetted radius and maximum height of splat for three cases. Initial temperatures of the drop in three cases are 25$^o$C, 45$^o$C and 60$^o$C while initial substrate temperature is 25$^o$C in all cases. The drop diameter and impact velocity in each case are 3.2 mm and 0.4 m/s, respectively.

Figure 11a: Comparison of droplet shapes from experiments and simulations for a 3.2 mm water droplet impacting with a velocity of 0.40 m/s ($Re$ = 1418, $We$ = 7.1) from $t$ = 0 to 22 ms. Initial drop as well as substrate temperature is 25$^o$C (isothermal-impact).

Figure 11b: continuation of (a) for latter times

Figure 12: Comparison of experimental and numerical results for wetted radius ($R_w$) and maximum height ($Z_{max}$) of splat for a 3.2 mm water droplet impacting with a velocity of 0.40 m/s ($Re$ = 1418, $We$ = 7.1). Initial drop as well as substrate temperature is 25$^o$C (isothermal-impact).

Figure 13: Comparison of experimental and numerical results for wetted radius ($R_w$) and maximum height ($Z_{max}$) of splat for a 3.2 mm water droplet impacting with a velocity of 0.40 m/s ($Re$ = 2223 (initial value), $We$ = 7.1). Initial drop and substrate temperature are 45$^o$C and 25$^o$C, respectively.

Figure 14: Comparison of experimental and numerical results for wetted radius ($R_w$) and maximum height ($Z_{max}$) of splat for a 3.2 mm water droplet impacting with a velocity of 0.40 m/s ($Re$ = 2771 (initial value), $We$ = 7.1). Initial drop and substrate temperature are 60$^o$C and 25$^o$C, respectively.

## 11 Figures

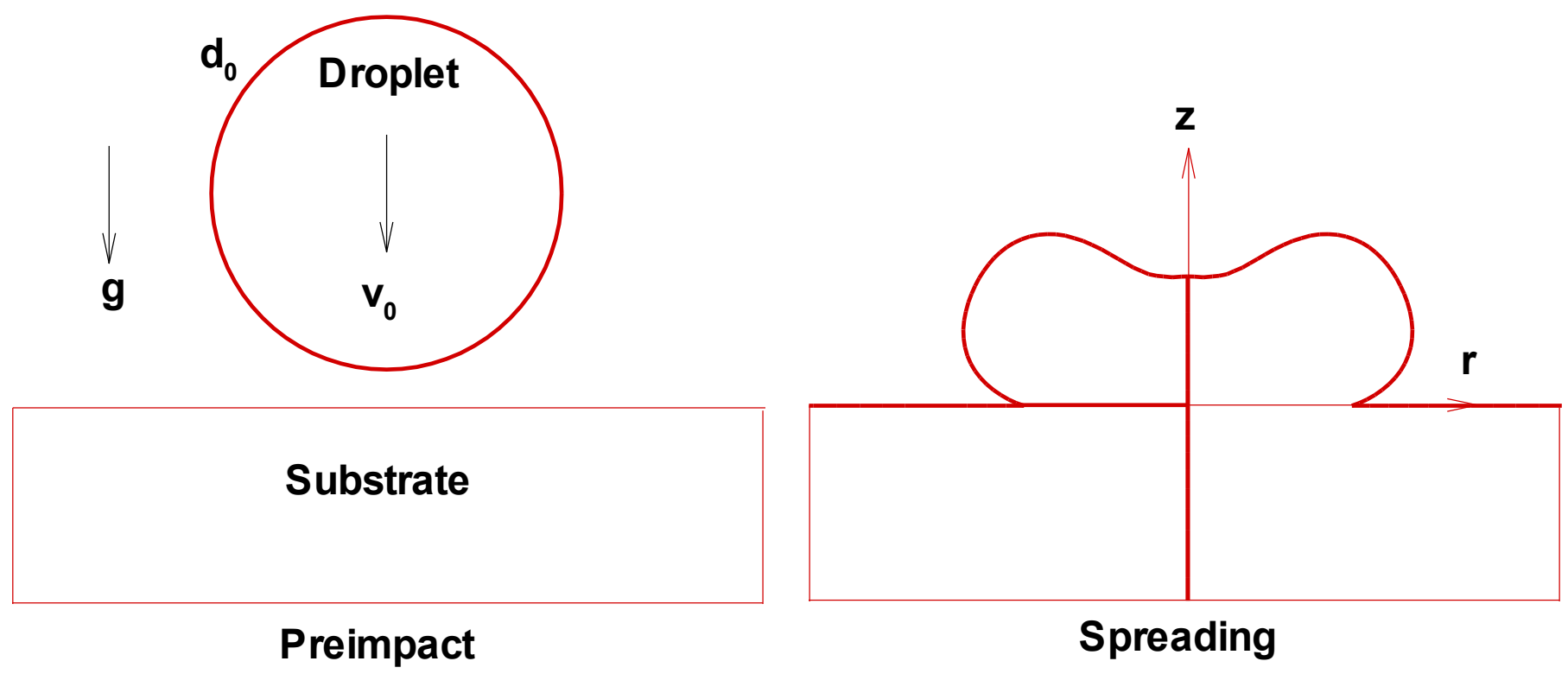

Figure 1: Definition of problem to be studied

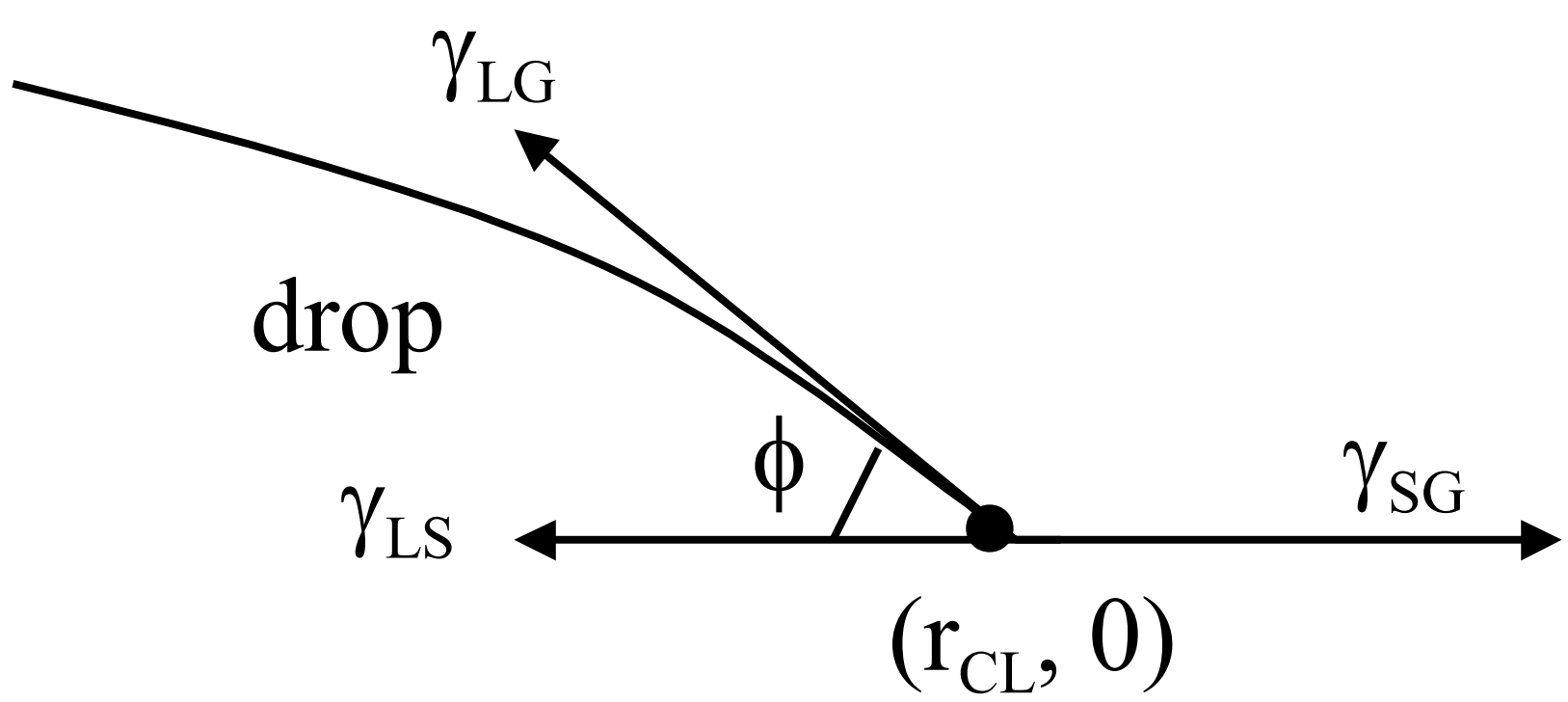

Figure 2: Surface forces at the wetting line

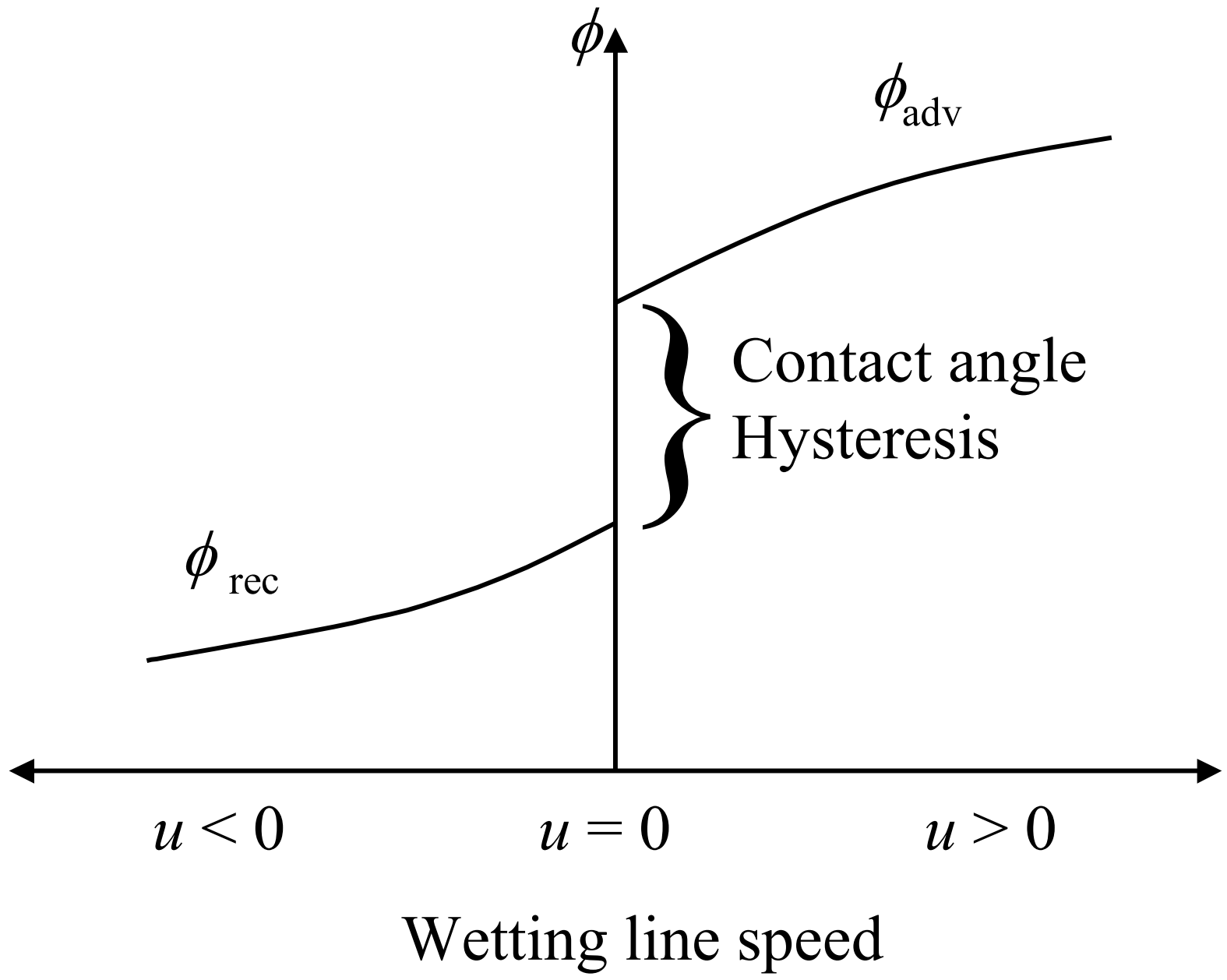

Figure 3: Relationship between the advancing and receding wetting angles, and the wetting line velocity. From [28].

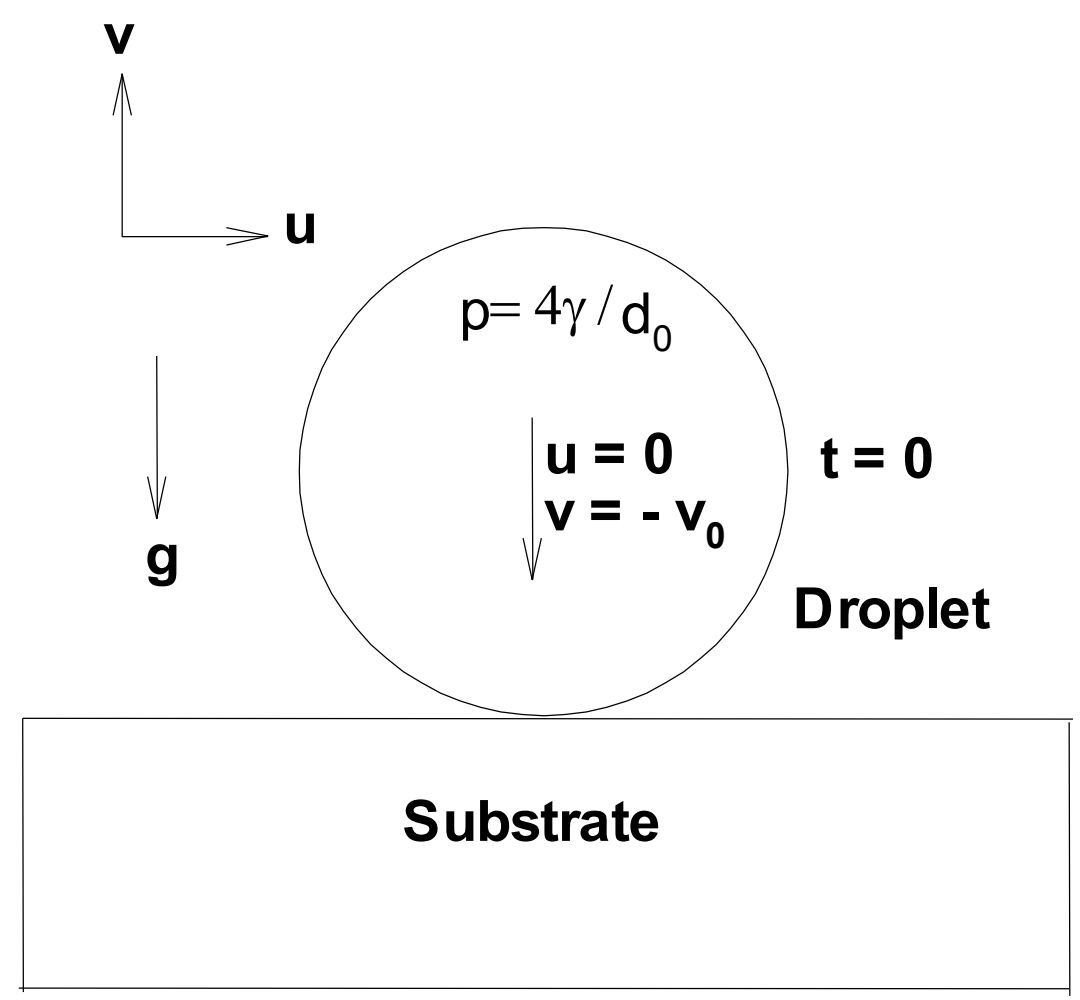

Figure 4: Initial conditions of the problem to be studied

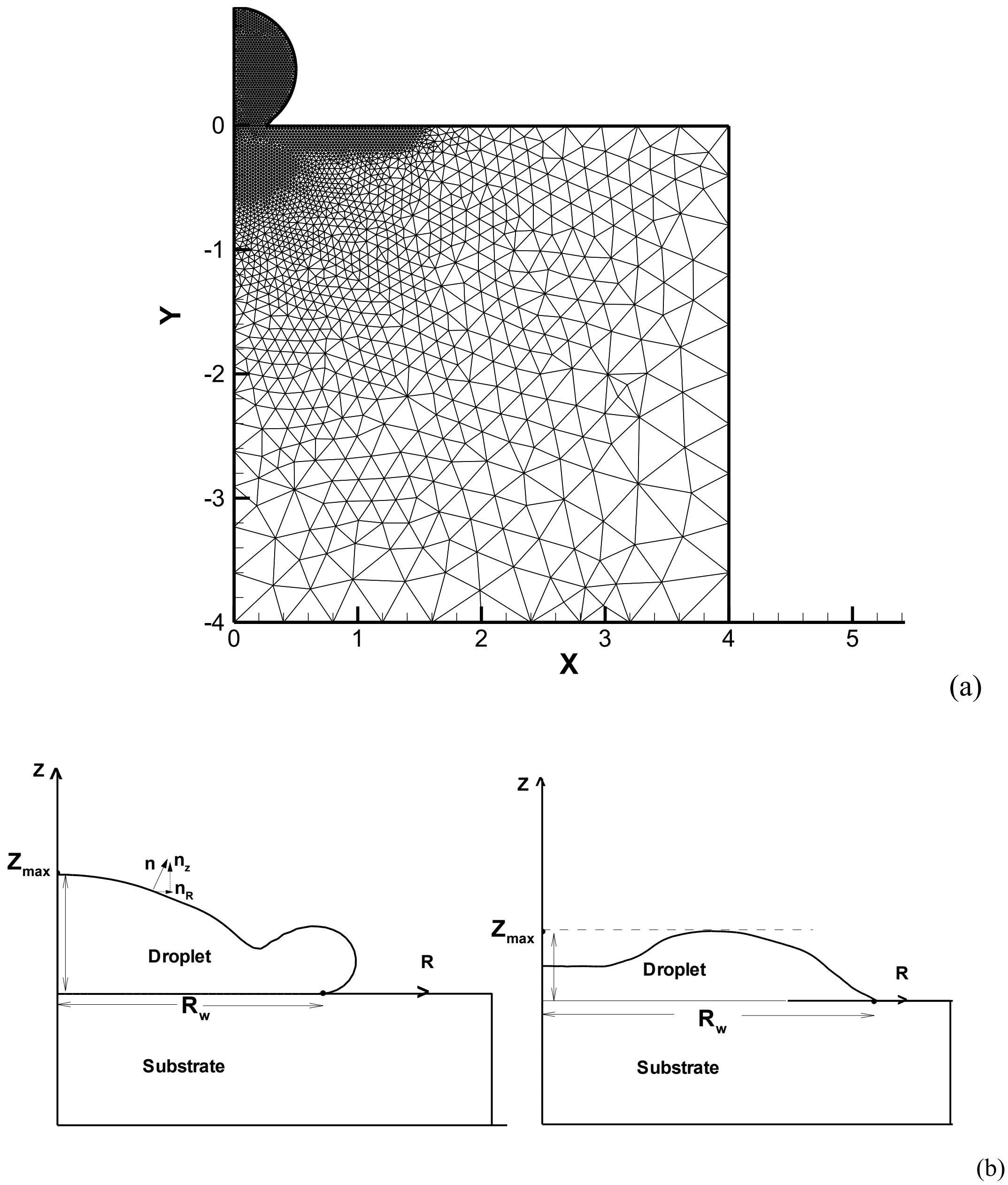

Figure 5 (a) A typical mesh in computational domain (b) Definition of the maximum height ($Z_{max}$) and wetted radius of droplet ($R_w$)

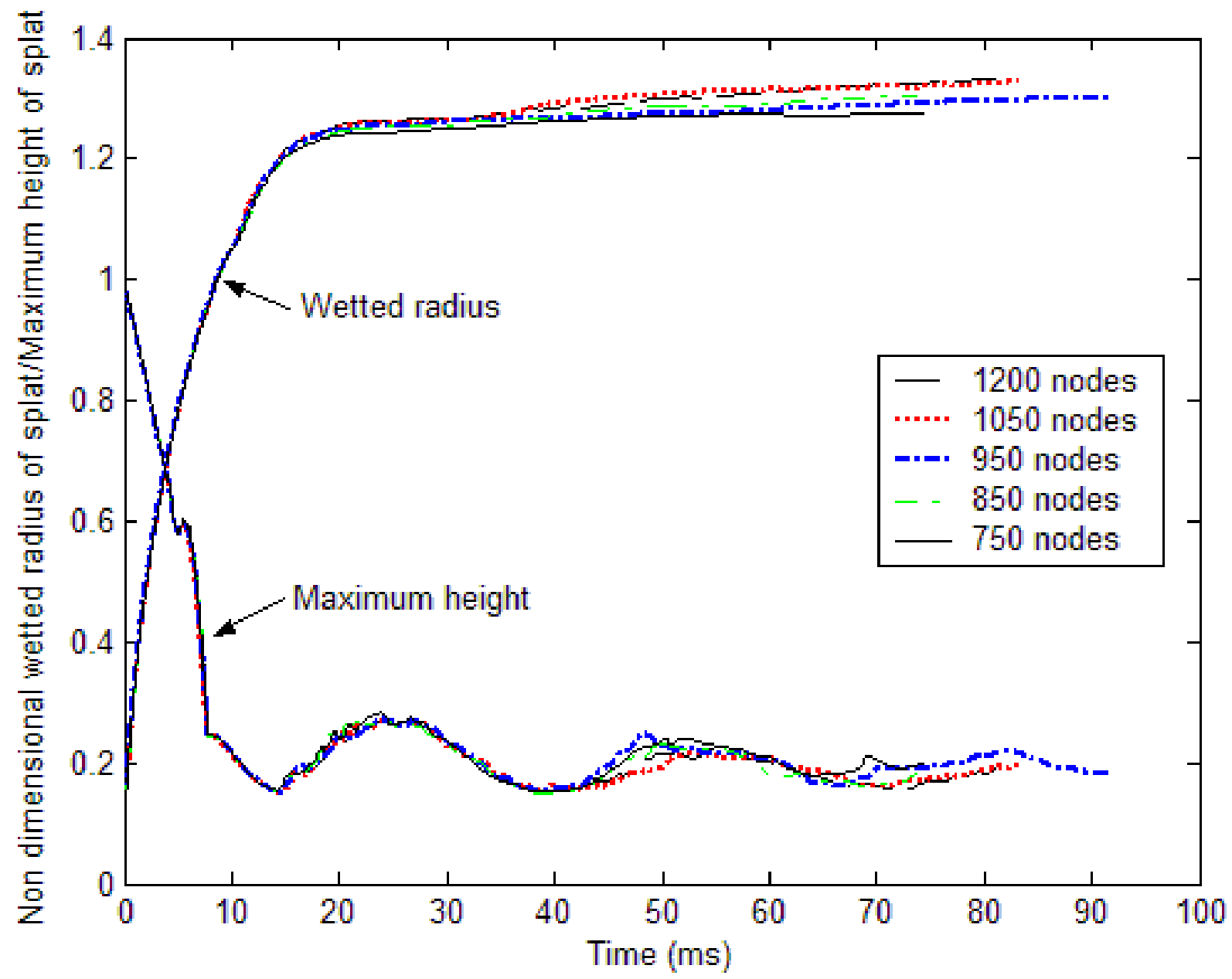

(a)

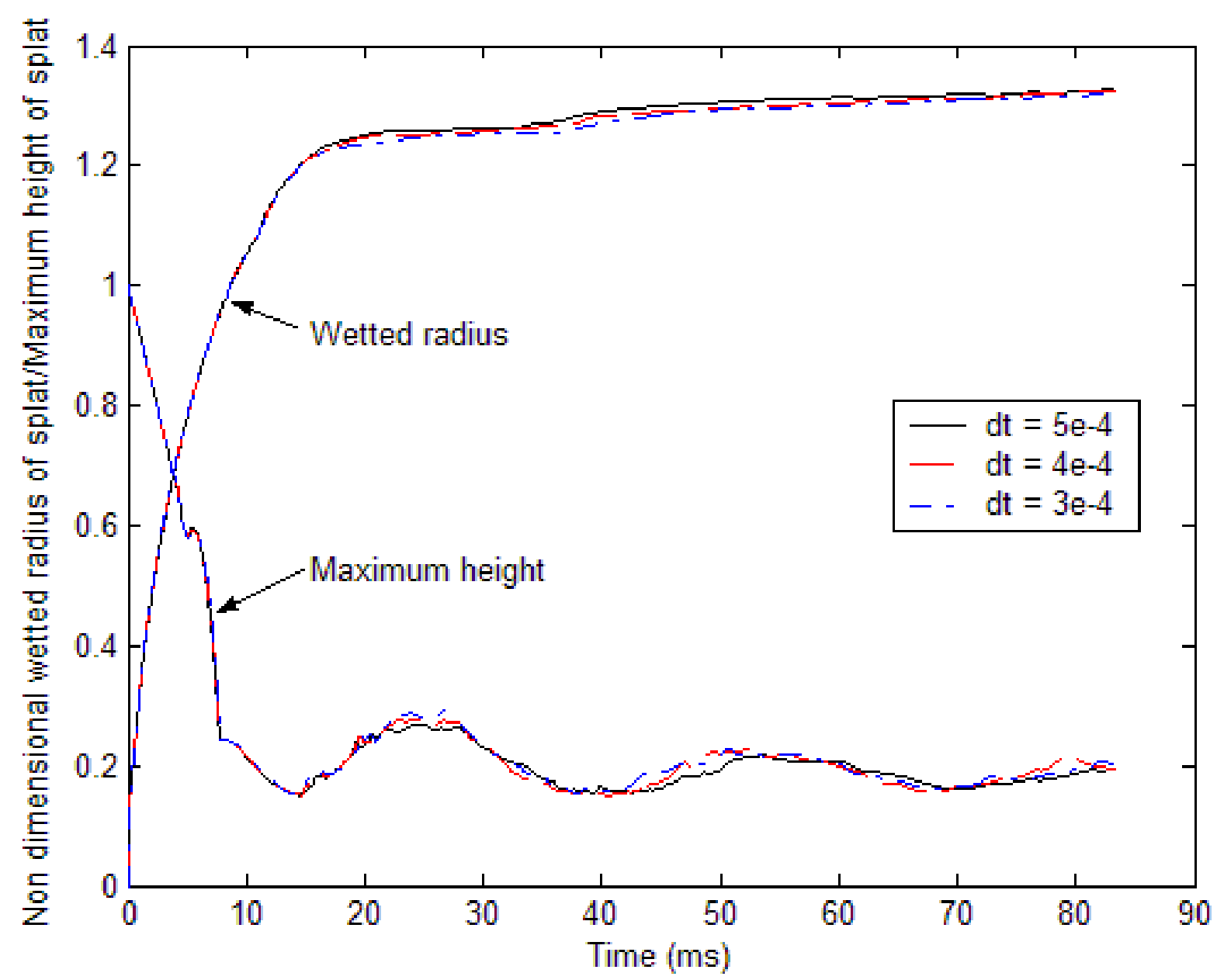

(b)

Figure 6: (a) Grid independence study: Variation of wetted radius and maximum height of droplet with time for different numbers of nodes in the droplet (b) Time-step independence study: Variation of wetted radius and maximum height of droplet with time for different time steps.

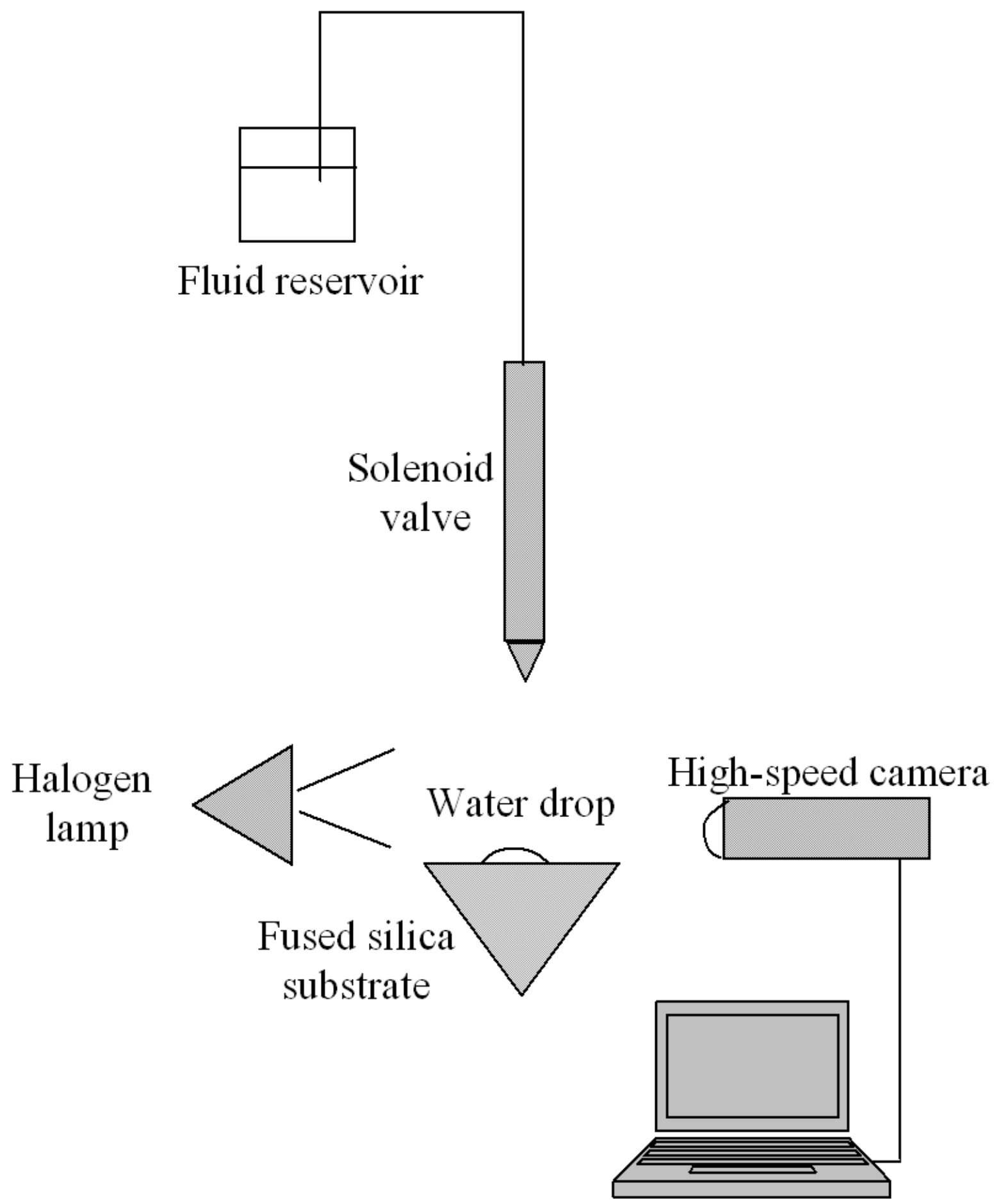

Figure 7: Experimental set-up

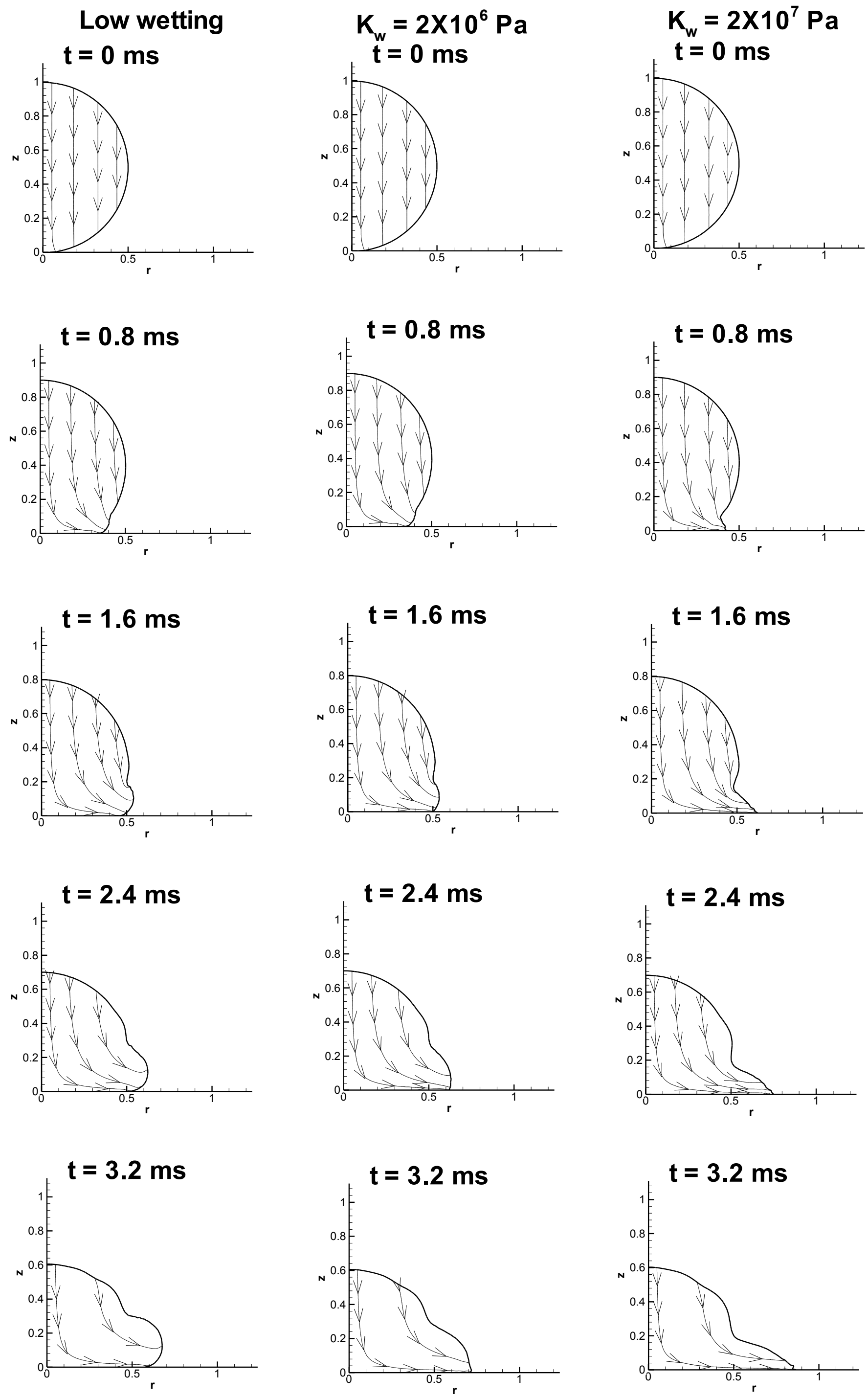

Figure 8a: Fluid dynamics during isothermal impact of a water drop on fused silica substrate (0 to 3.2 ms) for three different treatment of wetting (low wetting model, wetting with $K_w = 2\times10^6$ Pa and with $K_w = 2\times10^7$ Pa ). Diameter of drop and impact velocity are 3.2 mm and 0.40 m/s respectively (*Re* = 1418, *We* = 7.1) in each case. Initial temperature of drop as well as substrate is 25°C in each case (isothermal impact).

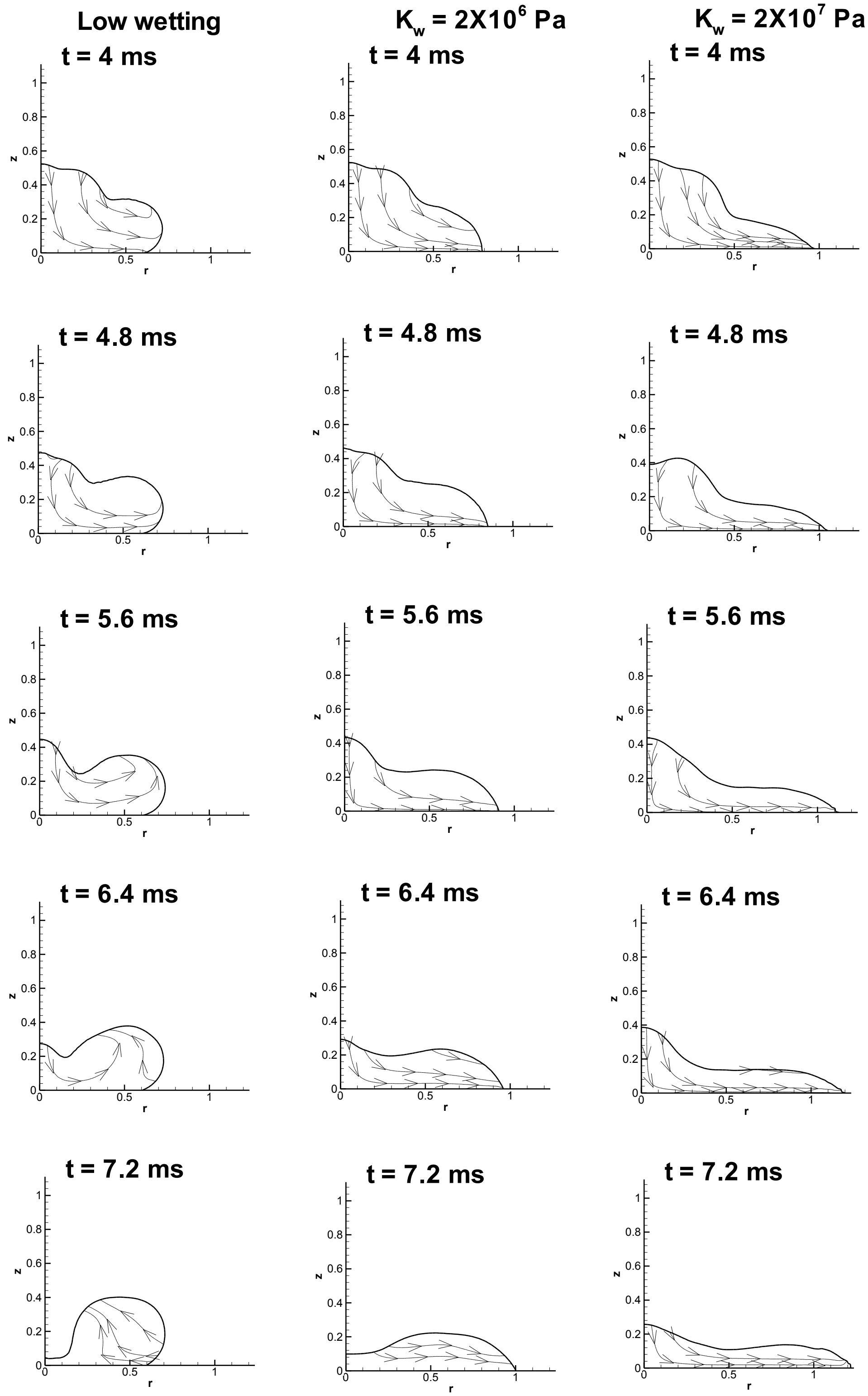

Figure 8b: continuation of (a) for latter times

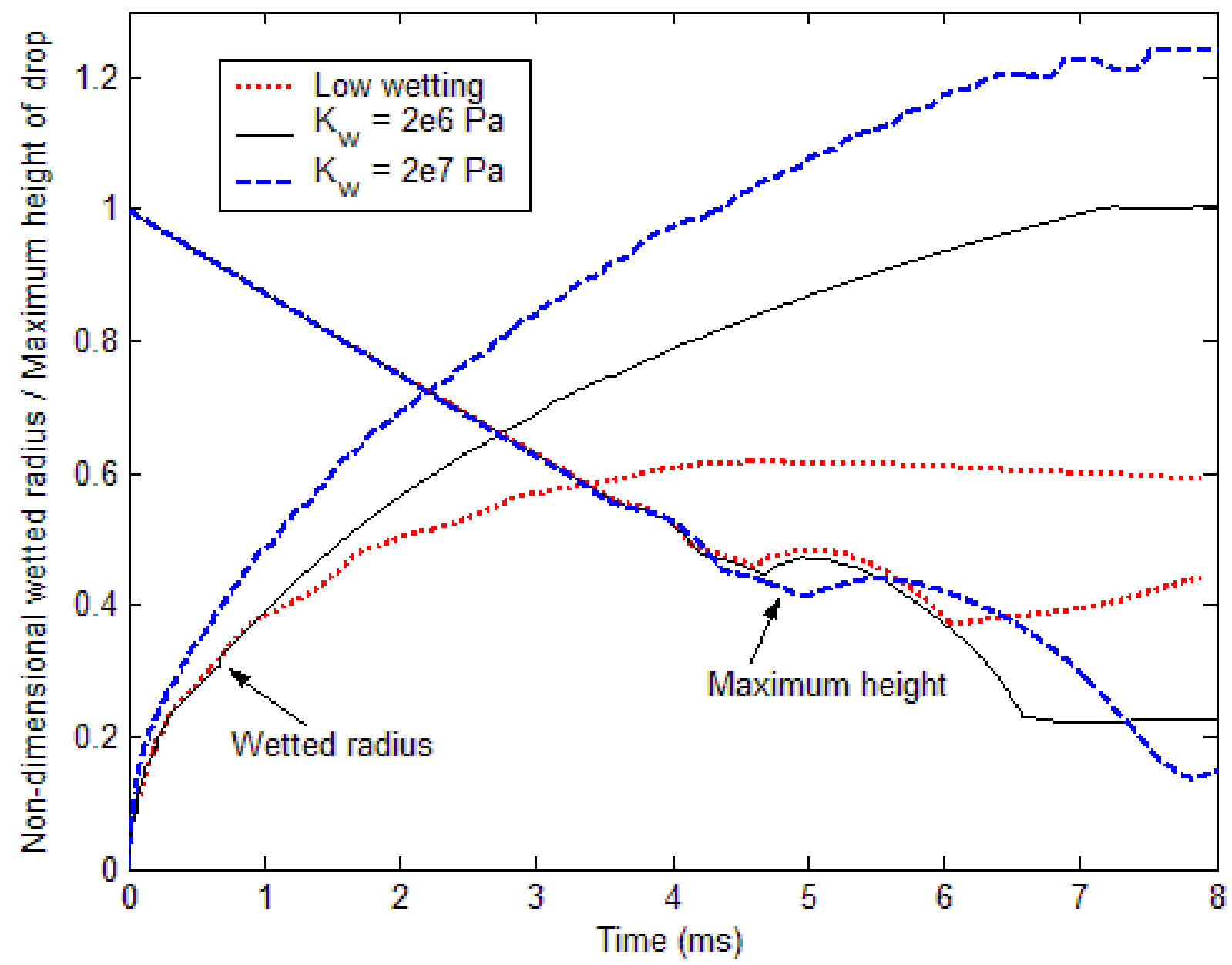

**(a)**

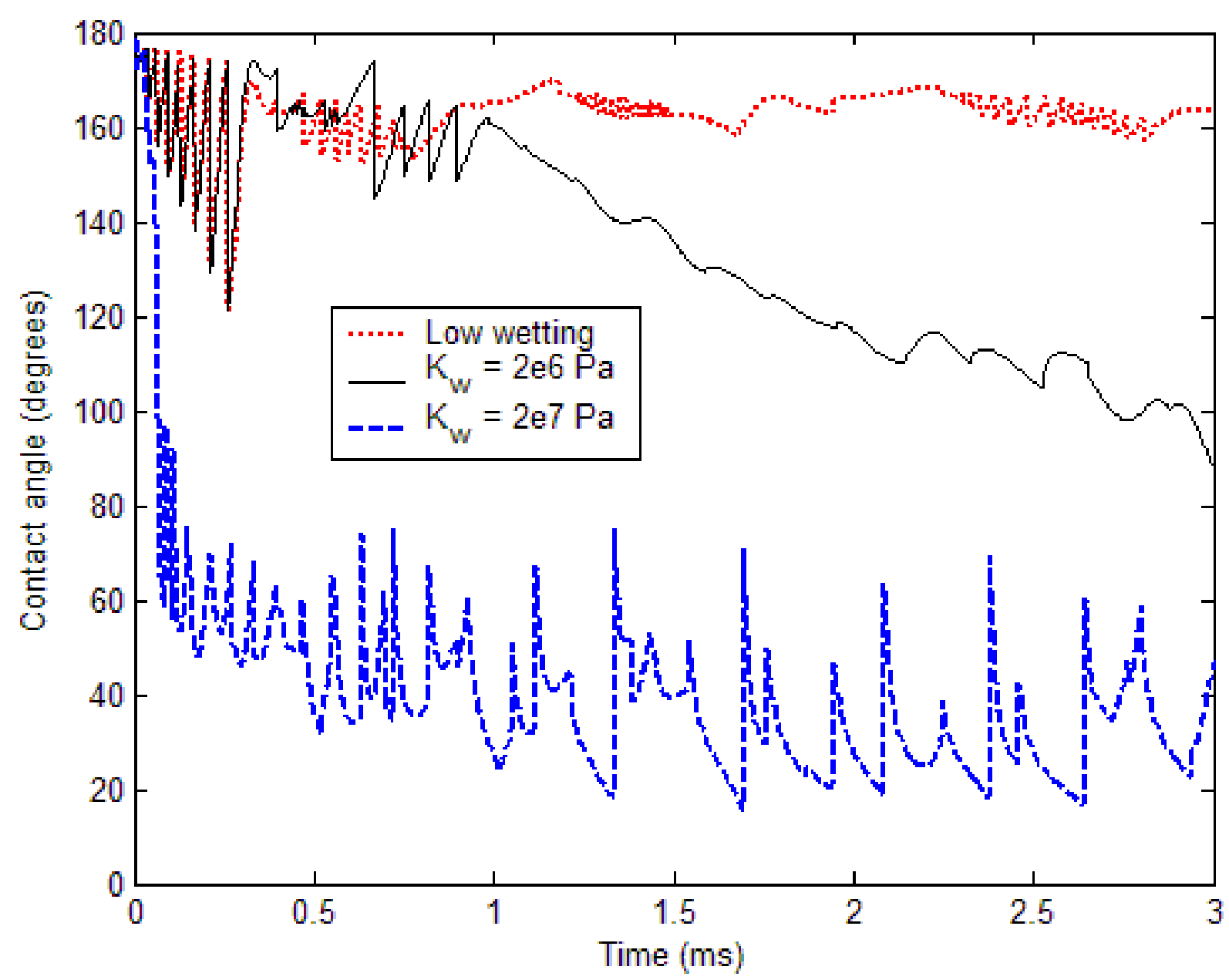

(b)

Figure 9: (a) Time evolution of dimensionless wetted radius and maximum height of splat for three different treatments of wetting (low wetting model, wetting with $K_w = 2\times10^6$ Pa and with $K_w = 2\times10^7$ Pa). The drop diameter and impact velocity in each case are 3.2 mm and 0.4 m/s, respectively. (b) Time evolution of contact angle for the three different treatments of wetting. Initial temperature of drop as well as substrate is 25°C in each case (isothermal impact).

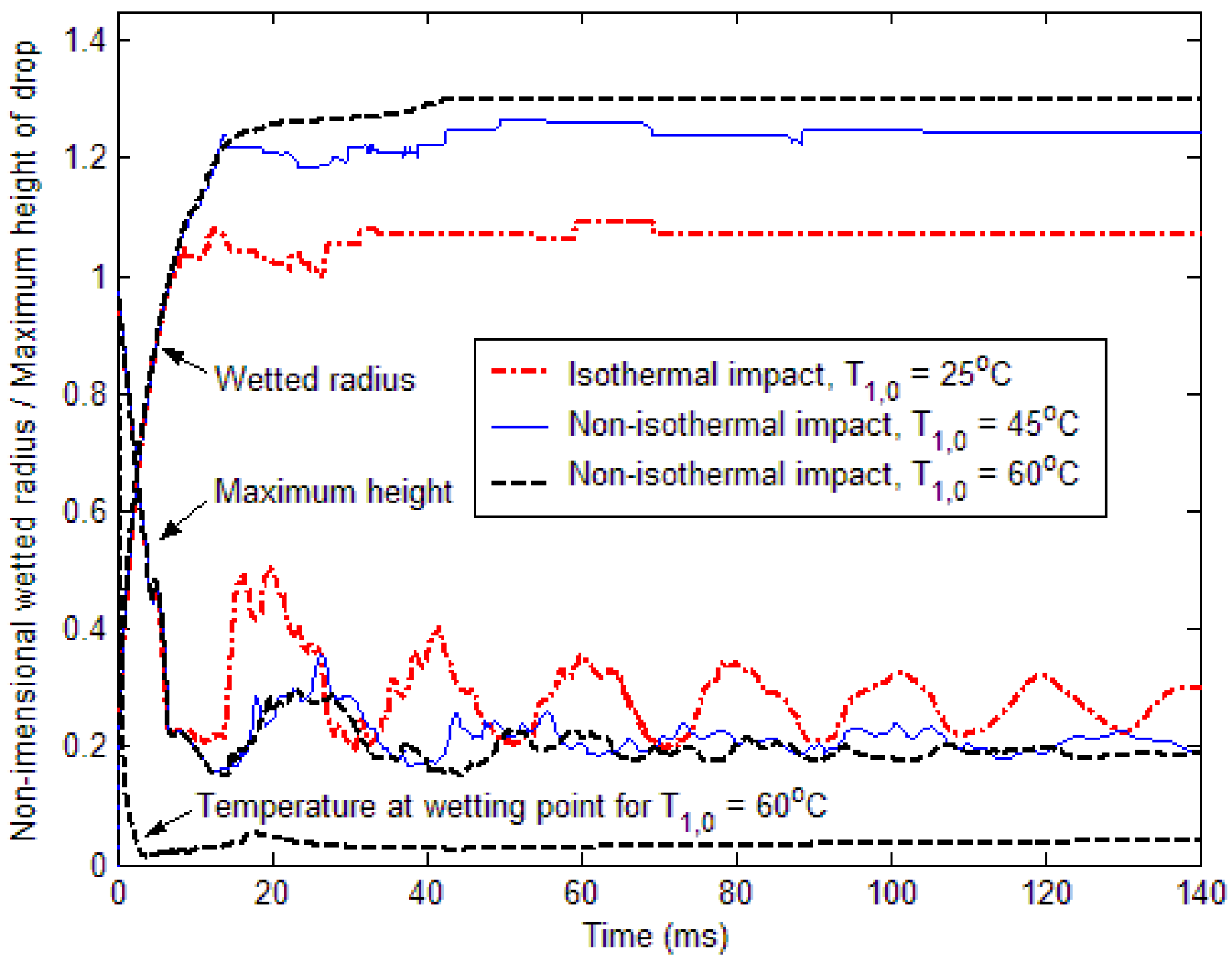

Figure 10: Time evolution of dimensionless wetted radius and maximum height of splat for three cases. Initial temperatures of the drop in three cases are 25°C, 45°C and 60°C while initial substrate temperature is 25°C in all cases. The drop diameter and impact velocity in each case are 3.2 mm and 0.4 m/s, respectively.

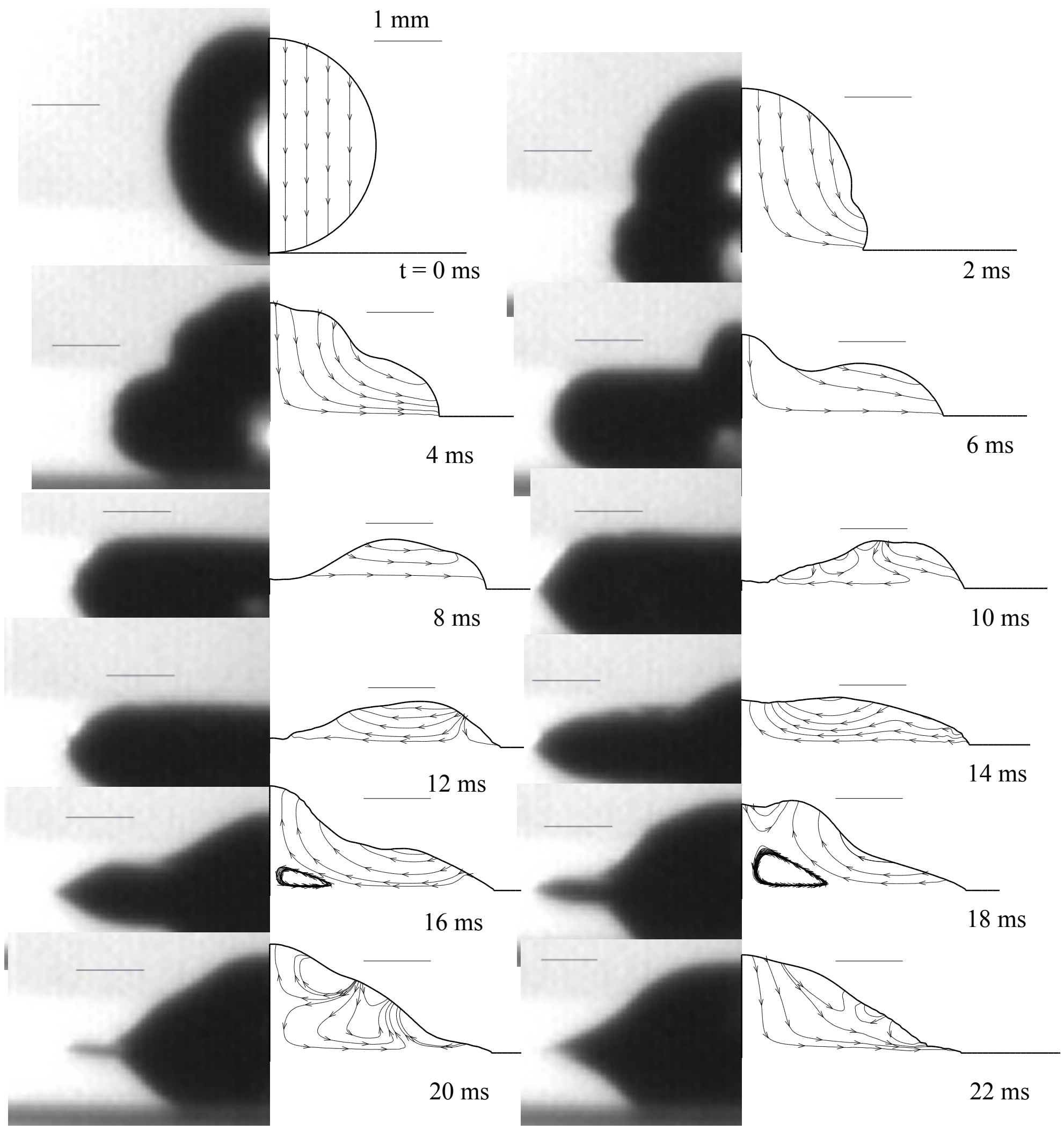

Figure 11a: Comparison of droplet shapes from experiments and simulations for a 3.2 mm water droplet impacting with a velocity of 0.40 m/s ($Re$ = 1418, $We$ = 7.1) from $t$ = 0 to 22 ms. Initial drop as well as substrate temperature is 25$^{o}$C (isothermal-impact).

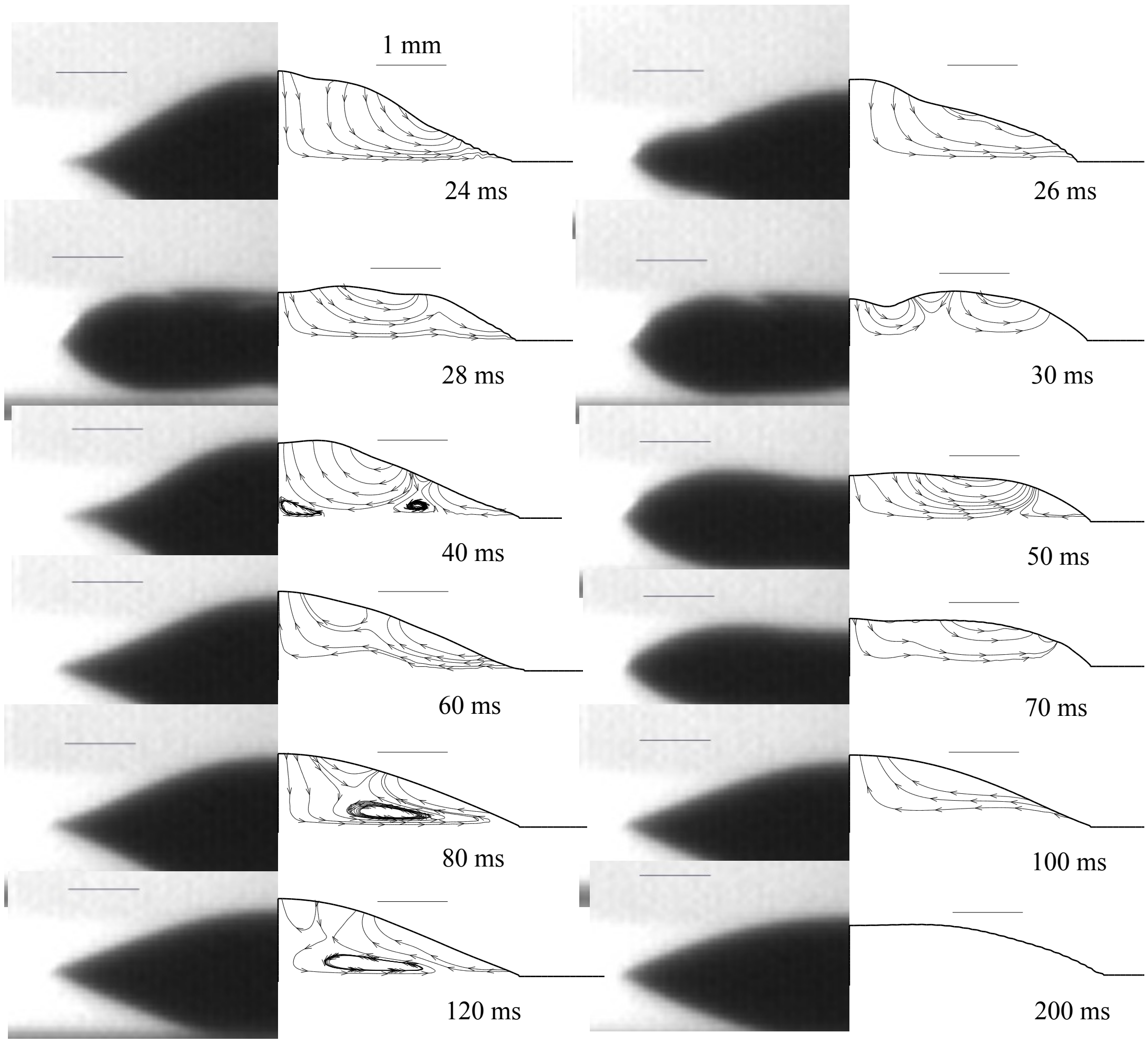

Figure 11b: continuation of (a) for latter times

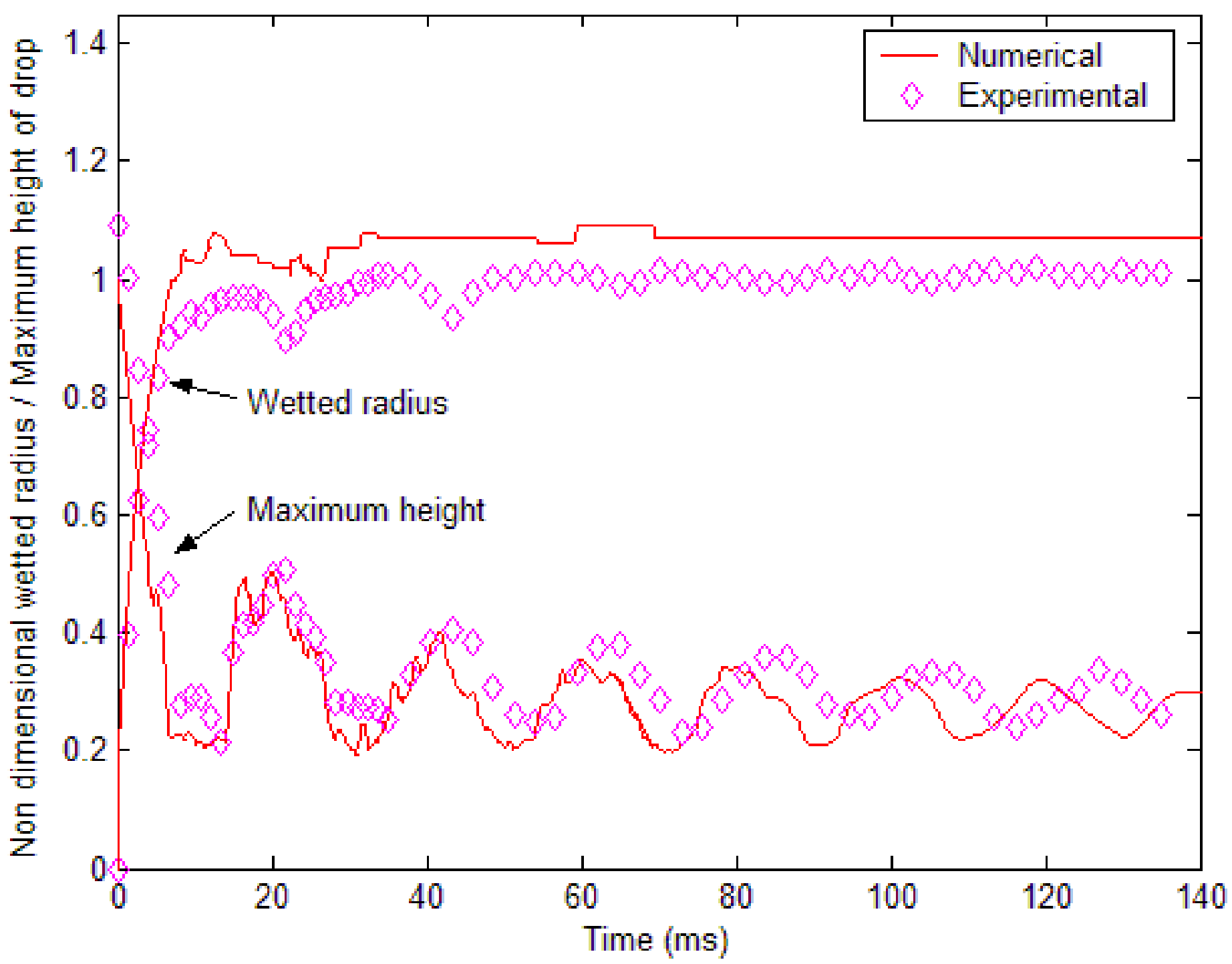

Figure 12: Comparison of experimental and numerical results for wetted radius ($R_w$) and maximum height ($Z_{max}$) of splat for a 3.2 mm water droplet impacting with a velocity of 0.40 m/s ($Re$ = 1418, $We$ = 7.1). Initial drop as well as substrate temperature is 25°C (isothermal-impact).

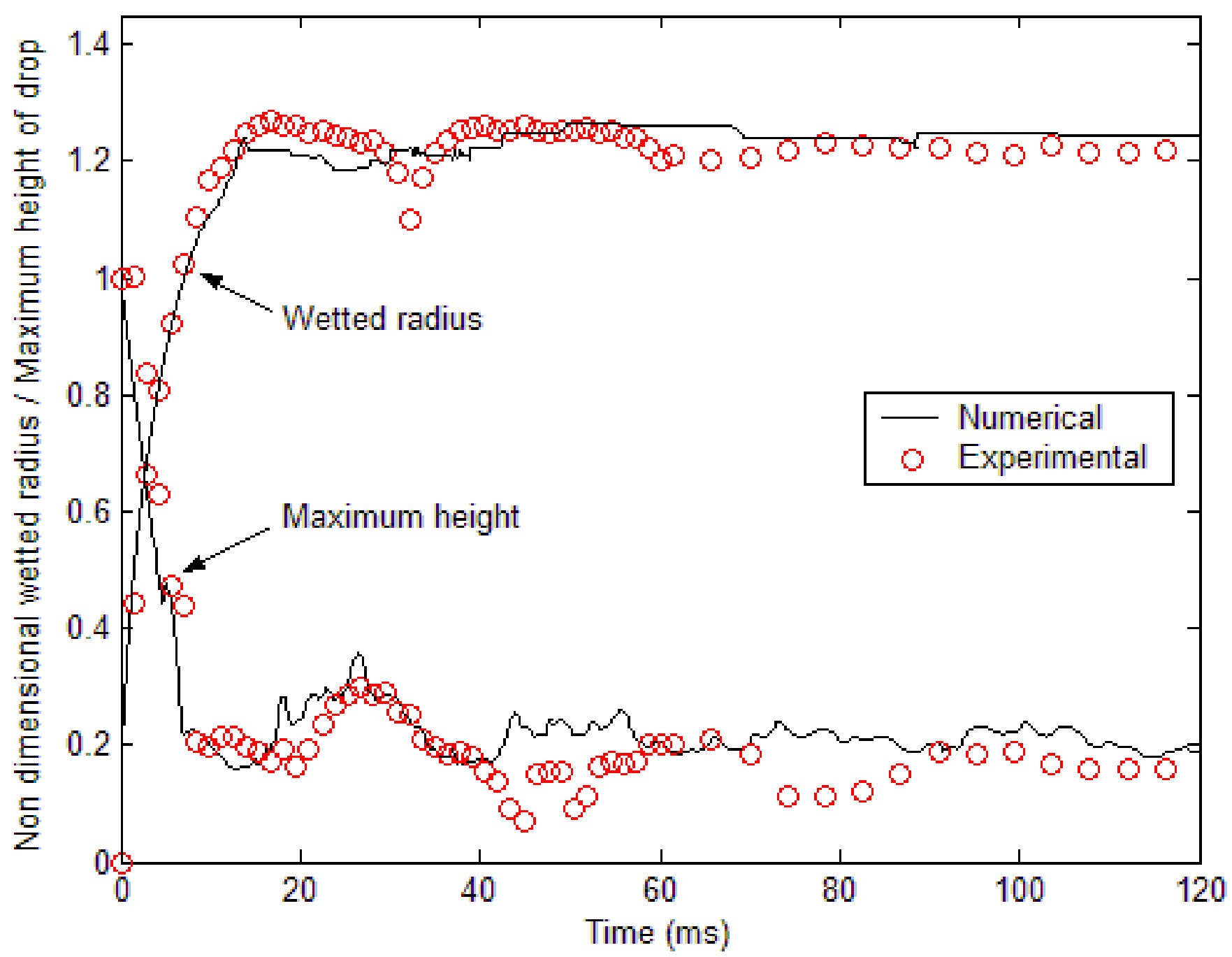

Figure 13: Comparison of experimental and numerical results for wetted radius ($R_w$) and maximum height ($Z_{max}$) of splat for a 3.2 mm water droplet impacting with a velocity of 0.40 m/s (*Re* = 2223 (initial value), *We* = 7.1). Initial drop and substrate temperature are 45°C and 25°C, respectively.

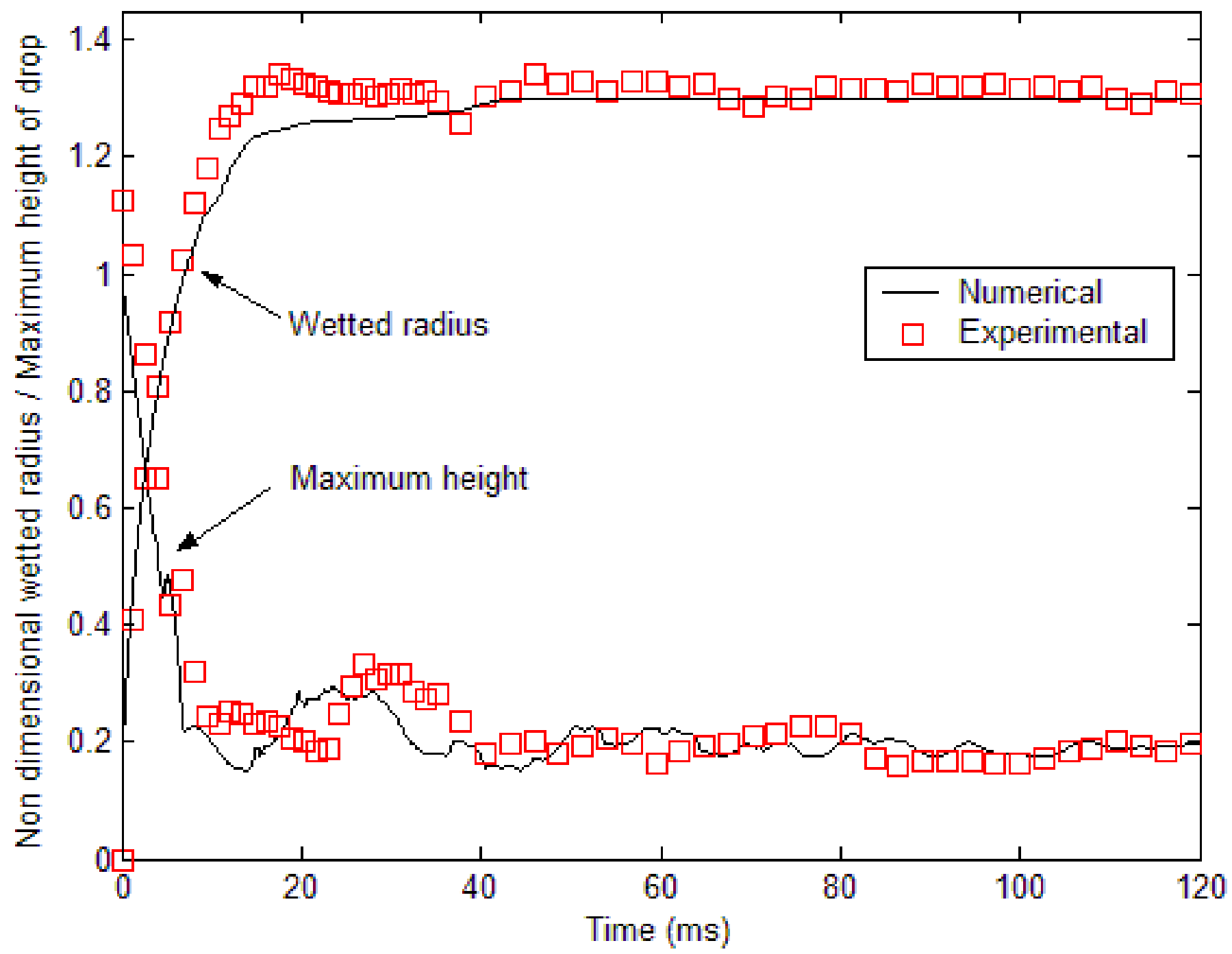

Figure 14: Comparison of experimental and numerical results for wetted radius ($R_w$) and maximum height ($Z_{max}$) of splat for a 3.2 mm water droplet impacting with a velocity of 0.40 m/s (*Re* = 2771 (initial value), *We* = 7.1). Initial drop and substrate temperature are 60°C and 25°C, respectively.